\documentclass[12pt]{amsart}
\usepackage[letterpaper,margin=1in]{geometry}
\date{\today}

\usepackage{amsmath}
\usepackage{amsfonts,amssymb}
\newcommand{\R}{\mathbb{R}}

\newcommand{\eq}[1]{eq.~(\ref{#1})}

\numberwithin{equation}{section}

\def\att#1{}

\newcommand{\Tr}{{\rm Tr}}
\newcommand{\ran}{{\rm ran}}
\newcommand{\ketbra}[2]{\vert #1\rangle\langle #2\vert}

\newtheorem{thm}{Theorem}[section]
\newtheorem{prop}[thm]{Proposition}

\newtheorem{lem}[thm]{Lemma}
\newtheorem{Definition}[thm]{Definition}

\newtheorem{Remark}[thm]{Remark}

\newcommand{\be}{\begin{equation}}
\newcommand{\ee}{\end{equation}}
\newcommand{\A}{\mathcal{A}} 

\renewcommand{\epsilon}{\varepsilon}

\begin{document}
\title{A Multi-Dimensional 
Lieb-Schultz-Mattis Theorem}
\author{Bruno Nachtergaele}
\address{Department of Mathematics\\ University of California at Davis\\
Davis CA 95616, USA}
\email{bxn@math.ucdavis.edu} 
\author{Robert Sims}
\address{Department of Mathematics\\ University of California at Davis\\
Davis CA 95616, USA}
\email{rjsims@math.ucdavis.edu}

\begin{abstract}
For a large class of finite-range quantum spin models with half-integer spins,
we prove that uniqueness of the ground state implies the existence of a 
low-lying excited state. For systems of linear size $L$, with arbitrary finite 
dimension, we obtain an upper bound on the excitation energy (i.e., the gap 
above the ground state) of the form $(C\log L)/L$. This result can be 
regarded as a multi-dimensional Lieb-Schultz-Mattis theorem \cite{Lieb1961}
and provides
a rigorous proof of the main result in \cite{hastings2004}.
\end{abstract}
\maketitle 
\renewcommand{\thefootnote}{$ $}
\footnotetext{Copyright \copyright\ 2006, 2007 by the authors. This article may be
reproduced in its entirety for non-commercial purposes.}
\setcounter{section}{0}


\section{Introduction and main result}\label{sec:intro}

\subsection{Introduction}\label{subsec:intro}
Ground state properties of Heisenberg-type antiferromagnets on a variety
of lattices are of great interest in condensed matter physics and 
material science.  Antiferromagnetic Heisenberg models are directly relevant 
for the low-temperature behavior of many materials, most notably the cuprates 
that exhibit high-$T_c$ superconductivity \cite{Manousakis1991}.

There are several general types of ground states that are known, 
or expected,  to occur in specific models: a disordered ground state or 
spin liquid, critical correlations (power law decay), dimerization 
(spin-Peierls states), columnar phases, incommensurate phases, and 
N\'eel order. More exotic phenomena such as chiral symmetry breaking have 
also been considered \cite{Sachdev1995,Sachdev2002}.

Which behavior occurs in a given model depends on the lattice, in particular
the dimension and whether or not the lattice is bipartite, on the type of spin
(integer versus half-integer) and, of course, also on the interactions. In this
paper we are considering  a class of half-integral spin models (or models where
the magnitude of at least some of the spins is half-integral). Our aim is to
prove a  generalization of the Lieb-Schultz-Mattis Theorem \cite{Lieb1961}.
Such a generalization was presented by Hastings in \cite{hastings2004} and a
substantial part of our proof is based on his work. Our main contribution is to
provide what we hope is a more transparent argument which in addition is
mathematically rigorous.

The well known theorem by Lieb and Mattis \cite{Lieb1962} implies,
among other things, that the ground state of the Heisenberg antiferromagnet
on a bipartite lattice with isomorphic sublattices, is non-degenerate.
For one-dimensional and quasi-one-dimensional systems of even length and 
with half-integral spin Affleck and Lieb \cite{affleck1986},
generalizing the original result by Lieb, Schultz, and Mattis \cite{Lieb1961},
proved that the gap in the spectrum above the ground state is bounded
above by {\em constant}$/L$. A vanishing gap can be expected to lead
to a gapless continuous spectrum above the ground state in the thermodynamic 
limit. Such an excitation spectrum is generically associated with power-law
(as opposed to exponential) decay of correlations.
Aizenman and Nachtergaele proved for the spin-1/2 antiferromagnetic chain
that if translation invariance is not broken (in particular, when the ground 
state is unqiue), the spin-spin correlation function can decay no faster than 
$1/r^3$ \cite{aizenman1994}. In other words, uniqueness of the ground state 
implies slow (power-law) decay of correlations. Recently, it was proved rigorously that a
non-vanishing spectral gap implies exponential decay of correlations 
\cite{nachtergaele2006a,hastings2006}. Therefore, non-exponential decay of
correlations implies the absence of a gap. In particular, the result by
Aizenman and Nachtergaele implies the absence of a gap in the infinite spin-1/2
antiferromagnetic chain if the translation invariance is not broken, e.g., if
the ground state is unique. This result can be generalized to an interesting
class of antiferromagnetic chains of half-integer spins \cite{Nachtergaele1994}.
The Lieb-Schultz-Mattis Theorem has also been extended to fermion systems
on the lattice \cite{yamanaka1997,tasaki2004}.
All these results are for one-dimensional systems. The bulk of the
applications of the spin-1/2 Heisenberg antiferromagnet is in two-dimensional
physics and therefore, the rigorous proof we provide here, based in part on 
ideas of Hastings \cite{hastings2004}, should be of considerable interest as 
it is applicable to higher-dimensional models.

The most common argument employed to bound a spectral gap from above uses the
variational principle. Often, the variational state is a perturbation of the
ground state. The proofs in \cite{Lieb1961} and \cite{affleck1986} are of this
kind. However, since the ground state is not known, and no assumptions are made
about it except for its uniqueness, these proofs are not a variational
calculation in the usual sense. The variational states are defined by acting
with suitable local operators $A$ on the (unknown) ground state. 

For a finite volume Hamiltonian $H_L$ generated by a
potential $\Phi$ of the type we consider (see the paragraph including
(\ref{eq:defnphia}) and (\ref{eq:hamL}) in Section~\ref{sec:setup} for the relevant definitions),
and with a unique ground state,
it is straightforward to show that the gap above the ground state, 
$\gamma_L$, is bounded uniformly in $L$.
To see this, note that for any ground state vector $\Omega$ and 
for any site $x$, there exists a unitary on the state space of $x$ with vanishing expectation in the state $\Omega$, i.e., $U\Omega\perp\Omega$.
Since $\Omega$ is the unique ground state by assumption, 
$U\Omega$ is a variational state for the gap. Therefore, we have
the bound
\begin{equation}
\gamma_L \leq \langle \Omega,[H_L,U]\Omega\rangle
\leq 2\inf_x \sum_{X\ni x} \Vert \Phi(X)\Vert
\leq 2 ||| \Phi |||_1\,,
\label{eq:ugb}\end{equation}
which is uniform in the system size $L$. Here,
$||| \Phi |||_1$ is as defined in (\ref{eq:1normphi}).
See Section \ref{sec:zerounitary} for the proof that such a unitary
exists. 

In order to obtain a better bound on the energy of the first excited 
state one has to exploit the few properties assumed of the ground
state, such as its uniqueness and symmetries. Furthermore, one
must show that any proposed variational state has a sufficiently large component in the
orthogonal complement of the ground state. In Section~\ref{subsec:vs}, we
propose a variational state for finite systems of size $L$ and then
demonstrate the relevant estimates, as mentioned above, in Sections
\ref{sec:ee} and \ref{sec:orthest}. 
It is interesting to note that the energy estimate we obtain 
will itself contain the spectral  gap of the finite system
in such a way that assuming a large gap leads to an upper bound less than the
assumed gap. From this contradiction one can conclude an upper bound on the
finite-volume gaps.

Our results apply to a rather general class of models, which we will define
precisely in the next section. The application of our general result to 
spin-1/2 Hamiltonians with translation invariant (or periodic) isotropic 
finite-range spin-spin interactions on a $d$-dimensional lattice is easy to 
state. 
First, let $\Lambda_L=[1,L]\times V_L$ with $L$ even and with periodic boundary 
conditions in the $1$-direction, i.e., in the direction that is of even size.
It will be important that the number of spins in $V_L$, $\vert V_L\vert$, 
is odd, and satisfies $\vert V_L\vert \leq c L^{d-1}$, for some $d\geq 1$ 
and a suitable constant $c$. Assuming that the model defined on $\Lambda_L$ has a unique
ground state, we prove that the spectral gap $\gamma_L$ satisfies the bound 
\begin{equation} 
\gamma_L \leq C \frac{\log L}{L}\, ,
\label{thebound}\end{equation}
where $C$ depends on $d$ and the specifics of the interaction, but not on $L$.

Because of the presence of the factor $\log L$, the bound (\ref{thebound}) 
applied to one-dimensional models does not fully recover the original 
Lieb-Schultz-Mattis Theorem in \cite{Lieb1961} or the bound proved by Affleck 
and Lieb in \cite{affleck1986}. This indicates that in general our
bound is not optimal. Our proof uses in an essential way Lieb-Robinson bounds 
\cite{Lieb1972,nachtergaele2006a,hastings2006}, as does Hastings' argument in
\cite{hastings2004}, and the appearance of the factor $\log L$ seems to
be an inevitable consequence of this. In fact, it is known that the standard
Heisenberg antiferromagnets with spin $\geq 1$ on the two-dimensional
square lattice or with spin $\geq 1/2$ on $\mathbb{Z}^d$ with $d\geq 3$,
have N\'eel ordered ground states \cite{Dyson1978,Kennedy1988} and in that case
one can show that the gap is bounded by $C/L$ (see, e.g., 
\cite{Landau1981,Koma1994}).

%
%
%
%

\subsection{Setup and main result}\label{sec:setup}

The arguments we develop below can be applied to a rather general class of
quantum spin Hamiltonians defined on a large variety of lattices. We believe
it is useful to present them in a suitably general framework which applies
to many interesting models. Attempting to be as general as possible, however,
would lead us into a morass of impenetrable notation. Therefore, we have
limited the discussion of further generalizations to some brief comments
in Section \ref{sec:generalizations}.

We assume that the Hamiltonians describe interactions
between spins that are situated at the points of some underlying set
$\Lambda$. For simplicity, one may think of $\Lambda = \mathbb{Z}^d$, 
but we need only assume that the set $\Lambda$ has one direction of
translational invariance, which we will refer to as the 
$1$-direction. We assume that there is an increasing sequence of 
sets $\{ \Lambda_L \}_{L =1}^{\infty}$ which exhaust $\Lambda$ of the form 
$\Lambda_L =  [1,L] \times V_L$ where $|V_L| \leq c L^{d-1}$ for some $d
\geq 1$. Here each $x \in \Lambda_L$ can be written as $x = (n,v)$
where $n \in \{ 1, 2, \cdots, L\}$ and $v \in V_L$, and we will denote
by $(n,V_L)$ the set of all $x \in \Lambda_L$ of the form $x=(n,v)$ for
some $v \in V_L$. 

Estimates on the decay of correlations in the ground state and 
Lieb-Robinson bounds on the dynamics will play an important role
in the proof of the main result. Both are expressed in terms
of a distance function on $\Lambda$, which we will denote by
$d$. Often, $\Lambda$ has the structure of a connected graph
and $d(x,y)$ is the minimum number of edges in a path from $x$ to $y$.
In any case, we will assume that $d$ is a metric and
furthermore that there is a function $F:[0,\infty)\to (0,\infty)$
satisfying the following two conditions.

\medskip
\noindent {\em Condition F1:} $F$ is uniformly integrable over $\Lambda$ in
the sense that
\begin{equation} \label{eq:fint}
\| \, F \, \| \, := \, \sup_{x \in \Lambda} \sum_{y \in \Lambda}
F(d(x,y)) \, < \, \infty.
\end{equation}

\medskip
\noindent
{\em Condition F2:} $F$ satisfies
\begin{equation} \label{eq:intlat}
C(F) \, := \, \sup_{x,y \in \Lambda} \sum_{z \in \Lambda} \frac{F \left( d(x,z) \right) \, F \left( d(z,y)
\right)}{F \left( d(x,y) \right)} \, < \, \infty,
\end{equation} 
which means that the ``convolution'' of $F$ with itself is bounded
by a multiple of itself.

F1 and F2 are restricitve conditions only when $\Lambda$ is infinite, however,
for finite $\Lambda$, the constants $\Vert F\Vert$ and $C(F)$ will
be useful in our estimates. It is also important to note
that for any given set $\Lambda$ and function $F$ that
satisfies F1 and F2 above, we can define a one-parameter family of
functions, $F_\lambda$, $\lambda \geq 0$, by
\begin{equation}
F_{\lambda}(x) := e^{- \lambda x} \, F(x),
\end{equation}
and easily verify that $F1$ and $F2$ hold for $F_\lambda$, with 
$\| F_{\lambda} \| \leq \| F \|$ and $C_{\lambda}(F) \leq C(F)$.

As a concrete example, take $\Lambda = \mathbb{Z}^d$ and $d(x,y) =
|x-y|$. In this case, one may take the function 
$F(x) = (1+x)^{-d - \epsilon}$ for any $\epsilon >0$. Clearly,
(\ref{eq:fint}) is satisfied, and a short calculation demonstrates
that (\ref{eq:intlat}) holds with 
\begin{equation}
C(F) \, \leq \, 2^{d + \epsilon + 1} \, \sum_{n \in \mathbb{Z}^d}
\frac{1}{(1+|n|)^{d+ \epsilon}}.
\label{latticeF}\end{equation}

Each $x\in\Lambda$ is assigned a finite-dimensional Hilbert space 
$\mathcal{H}_x$. For any finite subset $X \subset \Lambda$, the Hilbert 
space associated with $X$ is the tensor product 
$\mathcal{H}_X = \bigotimes_{x \in X}\mathcal{H}_x$, and 
the set of corresponding observables supported in $X$ is denoted by
$\mathcal{A}_X=\mathcal{B}(\mathcal{H}_X)$, the bounded linear 
operators over $\mathcal{H}_X$. These local observables form an 
algebra, and with the natural embedding of
$\mathcal{A}_{X_1}$ in $\mathcal{A}_{X_2}$ for any $X_1 \subset X_2$, 
one can define the $C^*$-algebra of all observables, $\mathcal{A}$, as 
the norm completion of the union of all local observable algebras 
$\mathcal{A}_{X}$ for finite $X \subset\Lambda$. Since we have assumed that
$\Lambda_L$ is of the form $[l,r]\times V_L$ with $r-l=L-1$, we can define 
translation automorphisms $\tau_n$, for $n\in\mathbb{Z}$, which map
$\A_{(m,V_L)}$ into $\A_{(n+m,V_L)}$ for all $m\in \mathbb{Z}$.

An interaction for the system is a map $\Phi$ from the
finite subsets of $\Lambda$ to $\A$ such that for each
finite $X\subset \Lambda$, $\Phi(X)^*=\Phi(X)\in \A_X$.
For given $\Lambda$ and $F$, and any $\lambda\geq 0$, 
let $\mathcal{B}_\lambda(\Lambda)$ 
be the set of interactions that satisfy
\begin{equation} \label{eq:defnphia}
\| \Phi \|_{\lambda} \, := \, \sup_{x,y \in \Lambda}  \sum_{X \ni x,y} \frac{ \| \Phi(X) \|}{F_{\lambda} \left(
    d(x,y) \right)} \, < \, \infty.
\end{equation}
All interactions considered in this paper are assumed to belong to
$\mathcal{B}_\lambda(\Lambda)$ for some choice of $F$ and $\lambda>0$.
The constant $\Vert \Phi \Vert_\lambda$ will show up in many estimates.
The finite volume Hamiltonians are defined in terms of the interaction $\Phi$ 
in the usual way by
\begin{equation} \label{eq:hamL}
H_L = \sum_{X \subset \Lambda_L} \Phi(X) + \mbox{ boundary terms}\, .
\end{equation}
We will always assume periodic boundary conditions in the 1-direction and 
arbitrary boundary conditions in the other directions (i.e., any boundary
terms in the other directions are included in the definition of $\Phi$).

The condition that $\Vert \Phi\Vert_\lambda$ is finite is sufficient
to guarantee the existence of the dynamics in the thermodynamic
limit as a one-parameter group of automorphisms on $\A$. In particular
this means that the limits
\begin{equation}
\alpha^\Phi_t(A):=\lim_{L\to\infty}\alpha_t^{\Phi,L}(A)
:=\lim_{L\to\infty} e^{itH_L}Ae^{-itH_L}
\label{dynamics}\end{equation}
exist in norm for all $t\in\mathbb{R}$, and all observables $A\in \A_X$,
for any finite $X\subset\Lambda$. We will often suppress 
the $L$ or $\Phi$ dependence in the notation $\alpha^{\Phi,L}_t$.
See \cite{bratteli1997,Simon1993,nachtergaele2006b} for more details.

Next, we turn to a set of conditions that more specifically describe the 
class of models to which the Lieb-Schultz-Mattis Theorem may be applied.

\medskip
\noindent
{\em Condition LSM1:} We assume that the interaction is translation invariant
in at least one direction, which we will take to be the 1-direction. This means 
\begin{equation}
\Phi\left( X + e_1 \right) = \tau_1 \left( \Phi(X) \right),
\end{equation}
where, for any $X\subset \Lambda$, $X+e_1$ is translation of all points in $X$
by one unit in the 1-direction. We will consider finite systems with
Hamiltonians $H_L$ defined with periodic boundary conditions in the 1-direction.
For convenience of the presentation we will assume free boundary conditions
in the other directions but this is not crucial.
Since we have assumed periodicity in the 1-direction, we can implement the 
translation invariance for finite systems by a unitary $T\in\A_{\Lambda_L}$ 
such that $\Phi(X+e_1)=T^* \Phi(X) T$, for all $X\subset \Lambda_L$. Here $T$
depends on $L$, but we suppress this dependence in the notation.

\medskip
\noindent
{\em Condition LSM2:} The interactions are assumed to be of finite range
in the 1-direction, i.e., there exists $R>0$ (the range), such that
if $X \subset \Lambda$ and $X \ni x_i = (n_i, v_i)$ for $i = 1, 2$ with
$|n_1 - n_2| \geq R$, then $\Phi(X) = 0$. 

\medskip
\noindent
{\em Condition LSM3:} We assume rotation invariance about one axis. More
precisely, we assume that there is a hermitian matrix in every $\A_{\{x\}}$,
$x\in\Lambda$, which we will denote by $S^3_x$, with eigenvalues that are
either all integer or all half-integer (i.e. belonging to $\mathbb{Z} + 1/2$).
We also require that $\tau_m (S^3_x) = S^3_{x+m e_1}$.
Define, for $\theta\in\mathbb{R}$, the unitary $U(\theta)\in \A_{\Lambda_L}$
by
\begin{equation} \label{eq:defu}
U(\theta) = \bigotimes_{x \in \Lambda_L} e^{i\theta S^3_x}\, .
\end{equation} 
The interaction is taken to be rotation invariant in the sense that
for each finite $X \subset \Lambda$
\begin{equation}
U^*(\theta) \Phi(X) U(\theta) = \Phi(X) \ \
 \mbox{for all } \theta \in \mathbb{R}\, .
\end{equation}

\medskip
\noindent
{\em Condition LSM4:} We assume that the $S^3_x$ are uniformly bounded: there
exists $S$ such that $\Vert S^3_x\Vert\leq S$, for all $x\in\Lambda$. The following
condition, which we call {\em odd parity}, is crucial:
define the parity of $x$, $p_x$ to be $0$ if the eigenvalues of $S^3_x$
are integers, and $p_x=1/2$ if they are half-integers. We assume that 
$\sum_{v \in V_L} p_{(n,v)} \in \mathbb{Z}+1/2$, for all $n\in \mathbb{Z}$. 
The simplest and most important case where this is satisfied is when we have 
a spin 1/2 at each site, and $\vert V_L\vert$ is odd.

\medskip
\noindent
{\em Condition LSM5:} The ground state of $H_L$ is assumed non-degenerate.
This implies it is an eigenvector of the translation $T$ and rotations
$U(\theta)$. Without loss of generality we can assume that $1$ is the 
corresponding eigenvalue of $T$ (if the eigenvalue is $e^{i\phi}$, 
replace $T$ by $e^{-i\phi}T$). We also assume that the ground state
has eigenvalue $1$ for the rotations $U(\theta)$.

\medskip
\noindent
{\em Condition LSM6:} We assume that there are orthonormal
bases of the Hilbert spaces $\mathcal{H}_{\Lambda_L}$ with respect to which
$S^3_x$ and $\Phi(X)$ are {\em real}, for all $x\in\Lambda_L,
X\subset\Lambda_L$. This condition is only used in the proof of
Lemma \ref{lem:partialsE}. Therefore, this condition may be replaced
by the property proved in that lemma. 

We will also use the following quantities:
\begin{equation}
|||\Phi |||_1\, := \, \sup_{x \in \Lambda} \sum_{X \ni x} \| \Phi(X) \| \,
< \infty,
\label{eq:1normphi}\end{equation}
and
\begin{equation}
|||\Phi |||_2 \, := \,
\sup_{x \in \Lambda} \sum_{X \ni x} |X| \, \sum_{x^\prime \in X}
\| [S^3_{x^\prime},\Phi(X) ]\| \, < \infty. 
\end{equation}
It is not hard to show that the conditions F1 and F2 are sufficient to
imply that $|||\Phi |||_1$ and $|||\Phi |||_2$ are finite.

We can now state our main result.

\begin{thm}\label{thm:main}
Let $\gamma_L$ be the spectral gap, i.e., the difference between the 
lowest and next-lowest eigenvalue of the Hamiltonian $H_L$ of a model
satisfying conditions F1, F2, and LSM1-6. Then, there exists a constant 
$C$, depending only on properties of $\Lambda$ (such as the dimension), 
the constants $\Vert F\Vert $ and $C(F)$, and the interaction 
($\Vert \Phi\Vert_\lambda$, for some $\lambda >0$, $||| \Phi |||_1$, and $||| 
\Phi |||_2$), such that
\begin{equation}
\gamma_L \leq C \frac{\log L}{L}\, .
\label{gapineq}\end{equation}    
\end{thm}

%
%
%
%
%

\subsection{Structure of the proof}\label{sec:proof}

The simplest way to present the proof is as a proof by contradiction.
Under the assumption that there exists a sufficiently large constant $C>0$,
such that $\gamma_L$ {\em exceeds} $(C\log L)/L$ for large $L$, 
we will construct a state orthogonal to the ground state with
an energy difference that is boundable by a quantity that is 
{\em strictly less than} the assumed gap for sufficiently 
large $L$. Thus, the proof is in essence a variational argument.
The variational state is constructed as a perturbation of the ground state,
as the solution of the differential equation proposed by Hastings
\cite{hastings2004} with the ground state as initial condition
(see Section \ref{sec:construction}, in particular (\ref{eq:exsde}), for this equation). The important
idea is that this equation will lead to a state which resembles the
ground state of the Hamiltonian with twisted rather than periodic
boundary conditions (see Section \ref{sec:twists} for the definition of the
twists), at least in part of the system, say the left half. In the right
half the ground state will be left essentially unperturbed.
This state is defined in Section \ref{sec:construction}.

After the variational state has been defined, there are two
main steps in the proof: estimating its excitation energy and 
verifying that it is ``sufficiently orthogonal'' to the ground state.
In general, one may also have to consider the normalization of the
variational state, but in our case the differential equation defining 
it will be manifestly norm preserving. Hence, this is not an issue for 
our proof.

The main difficulty is that under the general assumptions we have
made, no explicit information about the ground state is available.
Its uniqueness, translation, and rotation invariance are the only
properties we can use. In combination with the general assumptions 
on the interactions and the assumption on the magnitude of the
spectral gap above the ground state, however, one can obtain
an upper bound on the decay of correlations of the ground state
in the 1-direction. The recently proved Lieb-Robinson bounds
\cite{nachtergaele2006a,hastings2006,nachtergaele2006b} will be
essential to show that the effects of the perturbations we define
in the left half of the system remain essentially localized there. 
This allows us to compare the energy of the variational state
with the ground state energy of a Hamiltonian, $H_{\theta,-\theta}$
introduced in (\ref{eq:twham}), which, instead of twisted boundary
conditions, has two twists that cancel each other. The twisted Hamiltonian
is unitarily equivalent to the original one and therefore has the
same ground state energy. We work out this argument in Section
\ref{sec:ee}. 
The result is
\begin{equation} \label{eq:enerest}
\vert\langle \psi_1, H_L\psi_1\rangle - E_0 \vert
\leq C L^{\nu} e^{-c\gamma_L L}\, \left( 1 + \emph{corrections} \right) ,
\end{equation}
where $\psi_1$ is the normalized variational state we construct, and
$E_0$ is the ground state energy. The dependence of both quantities
on $L$ is surpressed in the notation. $\nu$, $C$ and $c$ are positive 
constants that only depend on properties of the lattice and the
interactions. The {\em correction terms} appearing above, and also in
(\ref{eq:obd+c}) below, can be made
explicit by the estimates provided in Section~\ref{sec:basest}. They
depend on the quantity $\gamma_L L$ in such a way that assuming
there exists a constant $C>0$ for which $\gamma_L L \geq C$ for
sufficiently large $L$, they are uniformly bounded in $L$. Due to the nature
of our proof of Theorem~\ref{thm:main}, see below, we do not write these
additional terms out explicitly. 

For the orthogonality, our strategy is to show that $\psi_1$ is
almost an eigenvector of the translation $T$ with eigenvalue $-1$.
Since the ground state $\psi_0$ is an eigenvector of $T$ with 
eigenvalue 1, by assumption, this shows that $\psi_1$ is nearly
orthogonal to $\psi_0$. In Section \ref{sec:orthest}
we obtain a bound on their inner product of the form:
\begin{equation} \label{eq:obd+c}
\vert\langle\psi_1,\psi_0\rangle\vert
\leq C^\prime L^{\nu^\prime} e^{-c^\prime \gamma_L L}\, \left( 1 + \emph{corrections} \right) ,
\end{equation}
where $\nu^\prime$, $C^\prime$ and $c^\prime$ are positive 
constants similar to $\nu$, $C$ and $c$. 
    
The proof of Theorem \ref{thm:main} then easily follows.

\begin{proof}(of Theorem~\ref{thm:main}:)
Suppose that $\gamma_L L \geq C \log L $ with a sufficiently large 
constant $C$. In this case, the {\em correction terms} which appear in
the bounds (\ref{eq:enerest}) and (\ref{eq:obd+c}) above are negligible. 
It is easy to see then that one obtains a contradiction for $L$ large enough. 
\end{proof}

To help the reader see the forest through the trees we have tried
to streamline the estimates in Sections \ref{sec:ee} and
\ref{sec:orthest} by collecting some results of a more 
technical nature in Section \ref{sec:basest}.

%
%
%
%

\subsection{Examples}\label{sec:examples}

The conditions LSM1-6 we have imposed on the models are not
unreasonable. We will illustrate this by considering various 
antiferromagnetic Heisenberg models defined on $\Lambda_L = [1,L] \times V_L$,
where for each $L$, $V_L$ is a finite set. As before, at each $x\in\Lambda_L$,
we have a finite-dimensional spin with spin-matrices $S^i_x$, $i=1,2,3$,
and we consider Hamiltonians of the form
\begin{equation}
H=\sum_{x,y\in\Lambda_L, x\neq y} J(x,y) \vec{S}_x\cdot\vec{S}_y\, ,
\label{heisenberg}\end{equation}
where $J(x,y)\in\mathbb{R}$ are the coupling constants. 

If $V_L\subset \mathbb{Z}^{d-1}$, with $d\geq 1$, and such 
$\vert V_L\vert \leq c L^{d-1}$, for a suitable constant
$c$, which describes the case for $d$-dimensional systems defined on 
subsets of $\mathbb{Z}^d$, there exists a function $F$ satisfying Conditions 
F1 and F2 as we have indicated in the paragraph containing (\ref{latticeF}).
It is also easy to see  that if $V_L$ is a fixed finite set independent
of $L$, in which case the system is (quasi) one-dimensional, any function
$F$ that works for the one-dimensional lattice will suffice.
All the examples we discuss below will be of this type.

Certainly, there are still many Hamiltonians of the form (\ref{heisenberg})
that fail to satisfy all six conditions, but this is generally for a  good
reason. For example, without translation invariance in at least one direction
one can easily have a non-vanishing gap above the ground state.

Condition LSM2, finite-range, does not need to be satisfied
in the strict sense. Sufficiently rapidly decaying interactions 
could also be treated. For the present discussion, let's assume that
the model is translation invariant in the 1-direction, and that
the interactions are nearest neighbor in the 1-direction in the
sense that for any $x=(n_1,v)$ and $y=(n_2,u)$, with 
$\vert n_1 -n_2\vert > 1$, we have $J(x,y)=0$.

The rotation invariance about at least one axis imposed in Condition LSM3 is
essential for the type of result we prove. The models (\ref{heisenberg}) have
full rotation invariance, so they clearly satisfy this condition. Anisotropic
models of the XXZ type would still satisfy LSM3.

In order to satisfy LSM4, we have to assume a uniform bound on the size of the
spin. Clearly, all models with only one kind of spins or a periodic arrangement
of spin magnitudes satisfy this condition. Since we already assumed translation
invariance in the 1-direction, we can verify the  {\em odd parity} condition by
adding the magnitudes of all spins in the ``slice'' $(1,V_L)$. If we have only
spin 1/2's, e.g., we simply need that $\vert V_L\vert$ is odd. For the
one-dimensional chains of identical spins of magnitude $S$, the condition
requires that $S$ is half-integral. Haldane's Conjecture
\cite{Haldane1983a,Haldane1983b}  predicts that for integer values of $S$ there
exists a non-vanishing gap. There are examples of isotropic integer-spin chains
which satisfy all the other conditions and for which the existence of a
non-vanishing gap has been rigorously established, such as the AKLT chain
\cite{affleck1988}. For $p$-periodic spin chains with a repeating pattern of
spin magnitudes $S_1,\ldots,S_p$, LSM4 is satisfied if $S_1+\cdots +S_p$ is
half-integral. Similarly, for spin ladders LSM4 is satisfied if the total spin
in each rung is half-integral.

There is a large class of models for which the uniqueness of the ground state
demanded  by LSM5 follows from the Lieb-Mattis Theorem \cite{Lieb1962}. 
For Hamiltonians of the form (\ref{heisenberg}), a simple case
where the Lieb-Mattis Theorem applies is the following:
if $\Lambda_L$ is the union of two disjoint subsets $\Lambda_{L,A}$ and
$\Lambda_{L,B}$ of equal size, with $J(x,y)\leq 0$ whenever $x$ and
$y$ do not belong to the same subset, and sufficiently many $J(x,y)$ are
non-vanishing such that the graph formed by the edges with non-zero coupling
constants is connected. This is satisfied if $V_L\subset
\mathbb{Z}^{d-1}$ is connected and the Hamiltonian is the usual nearest
neighbor antiferromagnetic Heisenberg model.

All models of the form (\ref{heisenberg}) satisfy LSM6.

The above discussion demonstrates that there is a large variety of
models that satisfy all conditions of our main theorem. In particular,
all nearest-neighbor half-integer spin Heisenberg antiferromagnets 
defined on subsets $\Lambda_L=[1,2L]\times V_L$ of $d$-dimensional hypercubic 
lattice with $\vert V_L\vert$ odd and such that 
$\vert V_L\vert \leq cL^\alpha$, for some $\alpha\geq 0$
(it is natural but not necessary to assume $\alpha=d-1$),
have a unique ground state with a gap $\gamma_L$ above it satisfying
$\gamma_L\leq C(\log L)/L$, for some constant $C$.

%
%
%
%

\subsection{Generalizations}\label{sec:generalizations}

One can envision several generalizations of Theorem \ref{thm:main}.
An obvious one is to remove the condition that the interaction is 
strictly finite range in the 1-direction. It is not hard to see that
the arguments given in the following sections can be extended to 
long-range interactions with sufficiently fast decay.

One may wonder whether the assumption that $L$ is {\em even} is essential.
It is used in the proof of near orthogonality of the variational state,
which is based on investigating the behavior under translations of 
the state: the variational state is close to an eigenvector with eigenvalue
$-1$ of the translation operator $T$, whereas the ground state has eigenvalue
$1$. Our proof of this fact assumes that the ground state is an eigenvector
of the rotations with eigenvalue $1$. For $L$ odd, our assumptions preclude
the existence of such an eigenvector. However, it seems plausible that for 
odd $L$ a slight modification of our proof will work to show that
the ground state and the variational excited state have opposite eigenvalues
for translations.

The main applications we think of are to SU(2)-invariant Hamiltonians
with antiferromagnetic interactions. 
Affleck and Lieb  \cite{affleck1986} pointed out that their proof
easily extends to a class of models with SU(N) symmetry.
There are no obstructions to generalizing our 
arguments to such models with SU(N) symmetry given by suitable
representations.

It may also be of interest to consider different topologies
of the underlying lattice and/or the twistings. Instead of 
cylindrical systems with periodic boundary conditions which can
be deformed by a twist, one could apply a similar strategy
to systems defined on a ball or a sphere. We do not
explore such possibilities here.

Another question we do not address in this paper is under what
circumstances the trial state we construct is actually a good
approximation of a low-lying eigenstate with energy close to
the first excited state, or even whether it is a state orthogonal
to the ground state and with energy bounded by $C(\log L)/L$.
We do not believe that statements of this kind hold under the
general conditions we impose. It is expected that in some
cases the true gap of the system is much smaller than the bound
we prove. This is of course not in contradiction with our result,
but under such circumstances our method does not provably construct 
a good variational state. There is no reason to assume that it
always would. 

%
%
%
%

\section{Construction of the variational state}\label{sec:construction}
 
\subsection{Twisted Hamiltonians}\label{sec:twists}\label{sec:twham}

The main motivation behind the construction of the variational
excited state is that it should resemble the ground state of the model
with twisted (as opposed to periodic) boundary conditions. Therefore,
we first describe some elementary properties of a family of perturbations
of the Hamiltonian, which we will call twisted Hamiltonians for reasons
that will become obvious.

Given an interaction $\Phi$ which satisfies the  general assumptions
outlined in Section~\ref{sec:setup}, we will now define a two parameter family
of ``twisted Hamiltonians'' to analyze. These Hamiltonians will be
defined on a finite volume $\Lambda_L = [1,L] \times V_L$, where 
$[1,L]$ is considered with periodic boundary conditions for some
even $L>4R$, where $R>0$ is the range of $\Phi$ in the 1-direction. 
Let $\Phi_L$ be the periodic extension of $\Phi$ restricted to $\Lambda_L$.
Recall that each point $x \in \Lambda_L$ can be written as 
$x=(n,v)$ where $n \in \{ 1, 2, \ldots, L \}$ and $v \in V_L$, 
and we will denote by $(n, V_L) = \{ x \in \Lambda_L \, : \, 
x=(n,v) \, \, \mbox{for some} \, \, v \in V_L \}$. For any $\theta \in
\mathbb{R}$ and $n \in \{1, 2, \ldots, L \}$, define the ``column''
rotations $U_n( \theta)$ by
\begin{equation}
U_n(\theta)=\bigotimes_{x\in (n,V_L)} e^{i\theta S^3_x}\, .
\label{eq:columntwist}\end{equation}
For $m \in \{1, 2, \ldots, L-1\}$, we will
denote by $V_m( \theta)$ the unitary given by
\begin{equation} \label{eq:unitary}
V_m( \theta) \, = \, \bigotimes_{m < n \leq L}U_n( \theta).
\end{equation}

The ``twisted Hamiltonians'' are defined to be perturbations of the
initial Hamiltonian with periodic boundary conditions
defined by
\begin{equation} \label{eq:iniham}
H \, = \, \sum_{X \subset \Lambda_L} \Phi_L(X)\, .
\end{equation}
The perturbations have the following form:
\begin{equation} \label{eq:pertatm}
H_{\theta}(m) \, := \, \sum_{X \subset \Lambda_L} 
V_m(\theta)^* \Phi(X) V_m(\theta) - \Phi(X).
\end{equation}
for $m\in [R, L-R]$ to avoid interactions across the seam created
by identifying $L+1$ with $1$. 
Note that here we use the original potential 
$\Phi$, and not its periodic extension $\Phi_L$. 
Clearly, if $X \subset \bigcup_{m<n \leq L}(n, V_L)$ or $X \subset
\bigcup_{1 \leq n <m}(n, V_L)$, then $V_m( \theta)^*
\Phi(X) V_m( \theta) - \Phi(X)$ will vanish by rotation invariance of
the interaction, and therefore only those interactions across the
column $(m, V_L)$ contribute in (\ref{eq:pertatm}). For $\theta,
\theta' \in \mathbb{R}$ and $m \in \{R, R+1, \ldots, L/2 -R \}$ fixed, we
define
\begin{equation} \label{eq:twham}
H_{\theta, \theta'} \, := \, H + H_{\theta}(m) + 
H_{\theta'}(m + L/2),
\end{equation}
to be a doubly twisted Hamiltonian. With $m$ fixed, we regard
$\Lambda_L$ as the disjoint union of two sets
\begin{equation}
\Lambda_L \, = \, \Lambda_L^{(W)} \cup \Lambda_L^{(S)},
\end{equation}
where $\Lambda_L^{(W)}$ consists of two windows, one about each column
at which a twist has been applied; namely
\begin{equation}
\Lambda_L^{(W)} \, := \, \Lambda_L^{(W)}(m) \cup \Lambda_L^{(W)}
\left( m + L/2 \right) \quad \mbox{and} \quad \Lambda_L^{(W)}(y) \,
:= \, \bigcup_{|n-y| \leq \frac{L}{4} - R}
  (n, V_L), 
\end{equation}
for $y \in \{ m, m+ L/2\}$. Moreover, $\Lambda_L^{(S)}$ comprises the remaining
strips in $\Lambda_L$.  Given this decomposition of the underlying  space, the
twisted Hamiltonian can be written as
\begin{equation} \label{eq:hamdecws}
H_{\theta, \theta'} \, = \, H_{\theta, \theta'}^{(W)} \, + \, H^{(S)},
\end{equation}
where
\begin{equation} \label{eq:hs}
H^{(S)} = \sum_{\stackrel{X \subset \Lambda_L:}{X \cap \Lambda_L^{(S)}
  \neq \emptyset}} \Phi_L(X),
\end{equation}
and $H_{\theta, \theta'}^{(W)}$ denotes the remaining terms in
$H_{\theta, \theta'}$ which, due to (\ref{eq:hs}), are supported
strictly within the windows.

There are a variety of useful symmetries the Hamiltonians $H_{\theta,
  \theta'}$, introduced in (\ref{eq:twham}), possess. With $m \in \{R,
R+1, \ldots, L/2 -R \}$ fixed as above, one may define
\begin{equation}
W( \phi) \, := \, \bigotimes_{m < n \leq m + L/2}U_n(- \phi),
\end{equation}
for any real $\phi$. See (\ref{eq:columntwist}) for the definition of the column
rotations $U_n$.
It is easy to check that for any
angles $\theta, \theta', \phi \in \mathbb{R}$, one has that
\begin{equation}
W^*( \phi) \, H_{\theta, \theta'} \, W( \phi) \, = \, H_{\theta -
  \phi, \theta' + \phi},
\label{equivHam}\end{equation}
due to the (term by term) rotation invariance of the
interactions. Given this relation, it is clear that along the path
$\theta' = - \theta$ the twisted Hamiltonian is
unitarily equivalent to the untwisted Hamiltonian, i.e.,
\begin{equation} 
W( \theta)^* \, H_{\theta, -\theta} \, W( \theta) \, = \, H_{0,0} \,
= \, H\, ,
\label{eq:unieq}\end{equation}
which, due to the periodic boundary conditions, is not true for general 
pairs $\theta, \theta^\prime$.

The untwisted Hamiltonian is translation invariant (in the
1-direction), and it is important that the twisted Hamiltonians
inherit a ``twisted'' translation invariance. Define
\begin{equation} \label{eq:twist}
T_{\theta, \theta^\prime} \, = \, T \, U_m( \theta) \, U_{m+L/2}(\theta^\prime).
\end{equation}
where $T$ is the unitary implementing the translation by $1$ 
in the $1$-direction. It is then straighforward to verify that
\begin{equation}
H_{\theta, \theta^\prime} \, = \, T_{\theta, \theta^\prime}^* \, H_{\theta,
\theta^\prime} \, T_{\theta, \theta^\prime}.
\end{equation}
Note that under the odd parity condition LSM4 we have
\begin{equation}
T_{2\pi,0}=-T\, ,
\end{equation}
which will be important in the proof of the almost orthogonality
of the trial state in Section \ref{sec:orthest}.

If we denote by $\psi_0$ the (unique) ground state of $H$, i.e., $H \psi_0 = E_0
\psi_0$, then by translation invariance, and specifically LSM5, we have
that $T \psi_0 = \psi_0$. Moreover, using the unitary equivalence (\ref{eq:unieq}), we see
that the ground state of the twisted Hamiltonian $H_{\theta, -\theta}$
satisfies $H_{\theta, -\theta} \psi_0( \theta, -\theta) = E_0( \theta,
-\theta) \psi_0( \theta, - \theta)$ with $E_0( \theta, -\theta) = E_0$
and $\psi_0( \theta, -\theta) = W( \theta) \psi_0$. Although the
twisted ground state $\psi_0( \theta, -\theta)$ is not translation
invariant, it does satisfy invariance with respect to the twisted
translations, i.e., $T_{\theta, -\theta} \psi_0( \theta, -
\theta) = \psi_0( \theta, -\theta)$. As a consequence, we have
the following simple but important property of $E_0$.

\begin{lem}\label{lem:partialsE}
Let $E_0(\theta, \theta^\prime)$ denote the ground state energy of
$H_{\theta,\theta^\prime}$.Then, the partial derivatives of $E_0$ vanish 
on the line $\theta^\prime=-\theta$:
\be \label{partialsE}
\partial_1 E_0(\theta,-\theta) =  \partial_2 E_0(\theta,-\theta) = 0.
\ee
\end{lem}
\begin{proof}
First, we note that $E_0$ is differentiable in its two variables
in a neighborhood of $(0,0)$ by the non-degeneracy condition LSM5.
By unitary equivalence $E_0$ is then differentiable in a neighborhood
of the line $(\theta,-\theta)$. For $\psi,\phi\in\mathbb{R}$, let 
$\mathcal{E}(\psi,\phi)=E_0(\psi-\phi,\psi+\phi)$ denote the ground state
energy of $H_{\psi-\phi,\psi+\phi}$. Due to the  unitary equivalence
\eq{equivHam}, $\mathcal{E}$ depends only on $\psi$.  Hence, $\partial_\phi
\mathcal{E}(\psi,\phi) = 0$, for all $\psi,\phi$. Under the additional
assumption that the interactions $\Phi(X)$ are {\em real} (LSM6),  we have that
$\overline{H_{\theta,\theta^\prime}}= H_{-\theta,-\theta^\prime}$, and
therefore $\mathcal{E}(\psi,0)=\mathcal{E}(-\psi,0)$. Hence, $\mathcal{E}$ is
an even function of $\psi$ and $\partial_\psi
\mathcal{E}(\psi,\phi)\vert_{\psi=0}=0$. Using these properties and the fact
that the partial derivatives of $E_0$ are  linear combinations of the partial
derivatives of $\mathcal{E}$, we find that both  partial derivatives of $E_0$
vanish on the line $\theta^\prime = -\theta$.
\end{proof}
%
%
%

\subsection{The variational state} \label{subsec:vs}

Our aim is to construct a state that resembles the ground state of 
$H_{\theta, - \theta}$ in a region surrounding those spins 
that were twisted by an angle of $\theta$, while it otherwise resembles 
the ground state of $H=H_{0,0}$.

{From} the unitary equivalence (\ref{eq:unieq}) we have that $E_0(\theta,-\theta)$
is independent of $\theta$, i.e., $\partial_{\theta} E_0(\theta,-\theta) =0$.
Moreover, the partial derivatives of $E_0$ vanish on the line
$(\theta,-\theta)$, as was proven in Lemma \ref{lem:partialsE}. 
This property, in general, allows one to derive an equation for 
the ground state.

Consider a differentiable one-parameter family of self-adjoint operators
$H(x)$, $x\in [a,b]\subset \mathbb{R}$, and let $E_0(x)$ denote the 
ground state energy of $H(x)$ with a differentiable family of
ground state eigenvectors $\psi_0(x)$. Suppose $\partial_x E_0(x) = 0$
for $x\in [a,b]$. Then, it is easy to see that 
$\psi_0(x) \perp (\partial_x H(x)) \psi_0(x)$, from which we obtain:
\begin{equation}
\partial_{x} \psi_0( x) \, = \, - \, \left[ H(
  x) - E_0(x) \right]^{-1} \, \partial_{x} H(x) \, \psi_0(
x).
\end{equation}
For any vector $\psi$,  this leads to
\begin{eqnarray} \label{eq:motcal}
\langle \psi, \partial_{x} \psi_0( x) \rangle & = &
- \int_{E_0(x) }^{ \infty} \, \frac{1}{E - E_0(x)} \, d \,
\left\langle \psi, P_E^{ x} \partial_{x}H( x) \psi_0(
  x) \right\rangle \\
& = & -  \int_{E_0(x) }^{ \infty} \, \int_0^{ \infty} \,
e^{-(E-E_0(x))t} \, dt \, d \,
\left\langle \psi, P_E^{ x} \partial_{x}H( x) \psi_0(
  x) \right\rangle \nonumber \\
& = & - \int_0^{ \infty} \,  \left\langle \psi, \alpha_{it}^{ x} \left(
  \partial_{x} H( x) \right) \psi_0( x) \right\rangle
\, dt, \nonumber 
\end{eqnarray}
where $P^x_E$ is the spectral resolution for $H(x)$ and 
$\alpha^{ x}_{it}$ is the imaginary-time evolution 
corresponding to the Hamiltonian. 

Motivated by this calculation, we introduce the family of operators
$B(A, H)$, where $H$ is a Hamiltonian for which the dynamics 
$\{ \alpha_{t} \mid t\in\mathbb{R} \}$ exists 
as a strongly continuous group of $*$-automorphisms and $A$ is any local
observable, defined by
\begin{equation} \label{eq:b0}
B(A, H) = - \int_0^{\infty} \alpha_{it}(A) dt,
\end{equation}
where $\alpha_{it}$ is the imaginary time evolution generated by $H$. 
For unbounded Hamiltonians $H$, it may not be obvious that $B(A, H)$ can be defined
on a dense domain. However, if $\psi$ is a ground state corresponding to the
Hamiltonian $H$, then $B(A, H) \psi$ exists. Moreover,
from (\ref{eq:motcal}), we conclude that
\begin{equation} \label{eq:gstatede}
\partial_{x} \psi_0(x) = B(x)  \psi_0(x),
\end{equation}
where $B(x) = B \left( \partial_{x}H(x), H(x) \right)$. 
Similarly, in the density matrix formalism, for 
\begin{equation} \label{eq:gsden}
\rho_0(x) \, := \, | \psi_0(x) \rangle \, \langle \psi_0(x) |,
\end{equation}
equation (\ref{eq:gstatede}) implies that 
\begin{equation} \label{eq:gsdende}
\partial_{x} \rho_0(x) \, = \, B(x) \rho_0(x) + \rho_0(x) B(x)^*
\end{equation}

We will define the proposed excited state $\psi$ as the solution of a
differential equation analogous to (\ref{eq:gstatede}). 
First, we need to introduce some further notation. Let $H$ be a
Hamiltonian for which the dynamics $\{ \alpha_t \}$ exists; finite
volume is sufficient. 
For any $a >0$, $t \in \mathbb{R} \setminus \{ 0 \}$, 
and local observable $A \in \A$, we may define 
\begin{equation} \label{eq:defa}
A_{a}(it, H) \, :=  \, \frac{1}{2 \pi i} e^{ - a t^2} \,
\int_{- \infty}^{ \infty} \, \alpha_s(A) \, \frac{e^{ - a s^2}}{s -
    i t} \, ds.
\end{equation}
In addition, for $T>0$ the quantity
\begin{equation}
 B_{a, T}(A , H) \, := \, - \int_0^T A_{a}(it, H) - A_a(it,
 H)^* dt,
\label{eq:batgen}\end{equation}
will play a crucial role. In Lemma~\ref{lem:gapbd} of 
Section~\ref{sec:basest}, we will show that when projected onto the
ground state of a gapped Hamiltonians $H$, the quantity $B_{a,T}(A, H)$ 
well approximates $B(A, H)$ for a judicious choice of parameters, e.g.,
$a=\gamma_L/L$ and $T=L/2$; 
we note that the observable $A$ must also satisfy
the constraint that its range is orthogonal to the ground state. 
With this in mind, consider the solution of the differential equation
introduced by Hastings in \cite{hastings2004}:
\begin{equation} \label{eq:exsde}
\partial_{\theta} \psi_{a, T}( \theta) = B_{a,T}(\theta) \, \psi_{ a , T}( \theta),
\end{equation}
where 
$B_{a,T}(\theta)=B_{a,T}(\partial_\theta H_{\theta,0}, H_{\theta, -\theta})$,
subject to the boundary condition $\psi_{a , T}(0) = \psi_0$.
Note that $B_{a,T}( \theta)$ is
anti-hermitian, and hence any $\psi_{a, T}( \theta)$ solving
(\ref{eq:exsde}) will have constant norm.

To be explicit, the proposed state 
$\psi_{a, T}(\theta)$ differs from the actual ground
state $\psi_0( \theta, - \theta)$, of the doubly twisted
Hamiltonian $H_{\theta, -\theta}$, in three essential ways. 
Compare (\ref{eq:b0}) in the case that $A= \partial_{\theta}H_{\theta, -\theta}$ 
and $H=H_{\theta,-\theta}$ with (\ref{eq:batgen}) given that 
$A =  \partial_{\theta} H_{\theta, 0}$ and $H= H_{\theta,-\theta}$. 
\newline i) We have introduced a cut-off at $T< \infty$. 
\newline ii) We have approximated the imaginary-time evolution of 
an observable $A$, $\alpha_{it}(A)$, by $ A_{a}(it, H) -A_{a}(it,H)^*$. 
\newline iii) We have replaced the observable 
$\partial_{\theta}H_{\theta, -\theta}$ with
$\partial_{\theta}H_{\theta,0}$. 

The modifications i) and ii) are of a technical nature, i.e., to make
the relevant quantities well-defined and amenable to estimation
(see Section \ref{sec:basest}). The motivation behind the third
replacement is an attempt to approximate the ground state of the singly twisted
Hamiltonian $H_{\theta, 0}$.

%
%
%
%

\section{Energy estimates} \label{sec:ee}

As is discussed in the introduction, the goal of this section is to
prove an estimate of the form
\begin{equation}
\vert\langle \psi_1, H_L\psi_1\rangle - E_0 \vert
\leq C L^{\nu} e^{-c\gamma_L L}\, ,
\label{eq:eest}\end{equation}
see (\ref{eq:enerest}) and Theorem~\ref{thm:energyest} below, for the proposed
variational state. Explicitly, we will take $\psi_1 = \psi_{a,T}( 2 \pi)$,
i.e., the solution of (\ref{eq:exsde}) evaluated at $\theta = 2 \pi$, with the
specific choice of parameters $a = \gamma_L / L$ and $T = L / 2$. Since the
operator $B_{a,T}( \theta)$, defined in (\ref{eq:exsde}), is
anti-hermitian, it is clear that $\psi_1$ remains normalized, and the bound
stated above demonstrates that if the gap is sufficiently large, $\gamma_L \geq
C \log(L) / L$,  then $\psi_1$ corresponds to a state with small
(depending on $C$) excitation energy. An estimate of the form (\ref{eq:eest}), \emph{with 
correction terms}, can be proven based on the general results in Section 
\ref{sec:basest}. In the proof of Theorem \ref{thm:main}, which is a proof by
contradiction, we will be assuming $\gamma_L \geq C \log(L) / L$. 
Therefore, we can assume here, without loss of generality, that there exists 
a constant $c>0$ such that $\gamma_L L \geq c$ for sufficiently large $L$. 
This assumption, which is not necessary, will simplify the presentation in 
Sections \ref{sec:ee} and \ref{sec:orthest}.

%
%
%
%

\subsection{Local estimates on the states}\label{sec:locee}

In this subsection we  prove a technical result which estimates, 
{\it uniformly in $\theta$}, the norm difference between the ground state
of $H_{\theta,-\theta}$ and the proposed state in the left half
of the system, more precisely in the window centered around the 
location, $(m,V_L)$ where the $\theta$-twist occurs. Since the restrictions
of the states to the half-systems are described by density matrices, it
is natural to use the trace norm for this estimate. Recall that for any bounded
operator $A$ on a Hilbert space $\mathcal{H}$, the trace norm is defined by
\begin{equation}
\Vert A \Vert_1 = \Tr \sqrt{A^* A}\, ,
\end{equation}
assuming this quantity is finite. Using the polar decomposition for
bounded linear operators, it is easy to see that, alternatively,
\begin{equation}
\Vert A \Vert_1 = \sup_{\stackrel{B\in\mathcal{B}(\mathcal{H}):}{ \Vert B\Vert =1}}
|\Tr AB|\, ,
\end{equation}

Recall that the density matrix corresponding to the ground state of the 
$H_{\theta, -\theta}$ Hamiltonian satisfies the equation
\begin{equation} \label{eq:gsev}
\partial_{\theta} \rho_0( \theta, - \theta) \, = \,
  B ( \theta) \, \rho_0(
  \theta, - \theta) \, + \, \rho_0( \theta, -\theta) \, B ( \theta)^*,
\end{equation}
compare with (\ref{eq:gsdende}), where we have used the notation
$B(\theta) = B ( A(\theta), H_{\theta, -\theta})$ for the
operator $B(A,H)$ as defined in (\ref{eq:b0}) and the observable 
$A(\theta) = \partial_{\theta}H_{\theta,-\theta}$. Note that
by construction $\rho_0(\theta,-\theta)$ remains normalized. 
We will often write $A(\theta) = A_1( \theta) - A_2( \theta)$ 
where, the observables $A_1( \theta) = \partial_{\theta} H_{\theta, 0}$
and  $A_2(\theta) = \partial_{\theta} H_{0, \theta}$ are
supported in the window about the twists of angle $\theta$ and
$-\theta$, respectively. Regarding $H_{\theta, \theta'}$ as a 
function of two variables, we may write $A_i( \theta) =
\partial_i H_{\theta, -\theta}$ for convenience. 
The notation $B_i( \theta) = B ( A_i(\theta), H_{\theta, -\theta})$
will also be useful. 

The proposed state is the solution of  
\begin{equation} \label{eq:exsev}
\partial_{\theta} \rho_{a,T}( \theta) = \left[
    \, B_{a,T} ( \theta), \rho_{a,T}( \theta) \, \right], 
\end{equation}
where the operator $B_{a,T}( \theta) = B_{a,T}( A_1(\theta), H_{\theta, -
    \theta})$ as in (\ref{eq:batgen}) with
observable $A_1(\theta) = \partial_1H_{\theta, - \theta}$.
The parametrization we choose is $a = \gamma_L /L$ and 
$T = L /2$. 
Since the operator $B_{a,T}(\theta)$ is anti-hermitian, the
solution $\rho_{a,T}( \theta)$ is a density matrix.  
We will denoted by $\mbox{Tr}_{m^c}[ \cdot ]$ the partial
trace over the Hilbert space corresponding to  $\Lambda_L^{(S)} \cup
\Lambda_L^{(W)}( m + L/2 )$. Note that the terms in the Hamiltonian
that have been twisted by an angle $\theta$ are supported in 
the complementary region $\Lambda_L^{(W)}(m)$.
Given a gap $\gamma_L>0$ above the ground state of the $H=H_{0,0}$
Hamiltonian, we will be able to estimate the
trace norm of the difference in the two states 
restricted to $\Lambda_L^{(W)}(m)$.
We will show this by estimating
$ \partial_{\theta} \, \Tr_{m^c} \left[ \rho_{a,T}( \theta) -
  \rho_0( \theta, -\theta) \right]$. 

\begin{thm} \label{thm:statedif}
As described in the introduction, we assume F1, F2, and LSM1-6.
If there exists a constant $c>0$ such that $\gamma_L L \geq c$ for
sufficiently large $L$ and we choose the 
parameters $a = \gamma_L / L$ and $T = L / 2$,
then there exists constants $C>0$ and $k>0$ so that
\begin{equation} \label{thm:ptbd}
\sup_{ \theta \in [0, 2\pi]} \left\Vert{\rm Tr}_{m^c} \left[ \rho_{a,T}( \theta) - \rho_0(
\theta, - \theta) \right] \right\Vert_1 \, \leq \, C \,
L^{2d} \, e^{ - k \gamma_L L}\,, 
\end{equation}
for $L$ large enough. Here $C$ and $k$ depend only on the interaction
$\Phi$ and the underlying set $\Lambda$.
\end{thm}

We note that the assumption concerning the existence of a constant
$c>0$ such that $\gamma_L L \geq c$ for sufficiently large $L$ is not
necessary. We impose it here for simplicity of presentation. Without
this additional assumption, one may prove an analogue of
(\ref{thm:ptbd}), which contains correction terms, by inserting the
bounds proven in Section~\ref{sec:basest} directly into the proof given below. 
Since we make this assumption, it is convenient to state a lemma
which compiles many of the technical results found in Section~\ref{sec:basest}
and applies them to the present set-up. For this, we need
two more definitions. Denote by $B_{a,T}^{(W)}( \theta)$ the operator defined by 
$B_{a,T} \left( A_1(\theta), H_{\theta, -\theta}^{(W)} \right)$ where the
Hamiltonian $H_{\theta, -\theta}^{(W)}$ is the full Hamiltonian
$H_{\theta, -\theta}$ restricted to the windows about the twists, 
see (\ref{eq:hamdecws}). Lastly, set $P_0^{\theta}$ to be the
projection onto the ground state $\psi_0( \theta, - \theta)$ of the
twisted Hamiltonian $H_{\theta, -\theta}$.

\begin{lem} \label{lem:bbd}
Assume F1, F2, and LSM1-6.
If there exists a constant $c>0$ such that $\gamma_L L \geq c$ for
sufficiently large $L$ and we choose the parameters $a = \gamma_L / L$ and $T = L / 2$,
then there exists constants $C>0$ and $k>0$ for which both
\begin{equation} \label{lem:nbd1}
\sup_{ \theta \in [0, 2 \pi]} \| B_{a,T}( \theta) - B_{a,T}^{(W)}(
\theta) \| \, \leq \, C \, L^{2d} \, e^{- k \gamma_L L}
\end{equation}
and
\begin{equation} \label{lem:nbd2}
\sup_{ \theta \in [0, 2 \pi]} \| \left( B_{a,T}( \theta) - B_1(
\theta)\right)P_0^{\theta}\| \, \leq \, C \, L^d \, e^{- k \gamma_L L}\,
\end{equation}
when $L$ is large enough.
\end{lem} 

\begin{proof}(of Lemma~\ref{lem:bbd})
Equation (\ref{lem:nbd1}) follows by combining Lemma~\ref{lem:lrbmt} and 
Remark~\ref{rem:decup}. Using Lemma~\ref{lem:gapbd} and Remark~\ref{rem:projest},
one obtains (\ref{lem:nbd2}).
\end{proof}

\begin{proof} (of Theorem~\ref{thm:statedif})
The proof of Theorem~\ref{thm:statedif} follows by deriving a uniform
bound on the $\theta$-derivative of the differences in these density
matrices. Specifically, the bound is in trace norm, and the uniformity
is with respect to $\theta \in [0, 2 \pi]$. 

Using (\ref{eq:gsev}), (\ref{eq:exsev}), and inserting the local
operator $B_{a,T}^{(W)}(\theta)$ for comparison, one may easily verify that
\begin{equation} \label{eq:drhodif}
\partial_{\theta} {\rm Tr}_{m^c} \left[ \, \rho_{a,T}( \theta) - \rho_0(
\theta, - \theta) \, \right] \, = \,  
{\rm Tr}_{m^c} \left( \left[ \,B_{a,T}^{(W)}(\theta), \,  \rho_{a,T}( \theta) - \rho_0(
\theta, - \theta) \, \right] \, \right) \,  + \, \sum_{i=1}^3 r_i( \theta),
\end{equation}
where the three remainder terms are given by 
\begin{equation} \label{eq:r1}
r_1(\theta) \,  : = \,  {\rm Tr}_{m^c} \left( \, \left[ \, B_{a, T} ( \theta)
     - B_{a, T}^{(W)}(\theta), \, \rho_{a, T}( \theta) \, \right] \, \right), 
\end{equation}
\begin{equation}
r_2 (\theta) \, := \, {\rm Tr}_{m^c} \left( \, \left[ \, B_{a,
      T}^{(W)} ( \theta), \, \rho_0(
  \theta, - \theta) \, \right] \, - \, \partial_1 \rho_0( \theta, -
\theta) \, \right),
\end{equation}
and
\begin{equation}
r_3 (\theta) \, := \, {\rm Tr}_{m^c} \left[ \partial_1 \rho_0( \theta, - \theta) \, - \,
\partial_{\theta} \rho_0 ( \theta, - \theta) \right].
\end{equation}
As $A_1(\theta)$ is supported near $(m, V_L)$ and $H_{\theta, - \theta}^{(W)}$ contains only those
interaction terms over sets $X \subset \Lambda_L^{(W)}$, it is clear
that $B_{a,T}^{(W)}( \theta)$ is contained in the algebra of 
local observables with support in $\Lambda_L^{(W)}(m)$; we will denote this
algebra by $\mathcal{A}(m)$.  Therefore,
\begin{equation}
{\rm Tr}_{m^c} \left( \left[
  B_{a,T}^{(W)}( \theta), \rho_{a,T}( \theta) - \rho_0(
\theta, - \theta) \right] \right) \, = \, \left[
  B_{a,T}^{(W)}( \theta), {\rm Tr}_{m^c} \left( \rho_{a,T}( \theta) - \rho_0(
\theta, - \theta) \right) \right].
\end{equation}
Since $ B_{a,T}^{(W)}( \theta)$ is anti-hermitian, we may apply norm preservation,
i.e. Theorem~\ref{thm:normp}, to (\ref{eq:drhodif}) and conclude that
\begin{equation}
\left\Vert \, {\rm Tr}_{m^c} \left[ \rho_{a,T}( \theta) - \rho_0(
\theta, - \theta) \right]\, \right\Vert_1 \, \leq \, \sum_{i=1}^3
\int_0^{\theta}  \Vert r_i(\theta') \Vert_1 \, d \theta'.
\end{equation}
We need only bound the trace norms of the remainder terms $r_i( \theta)$.

As $\rho_{a,T}( \theta)$ is a density matrix, in particular non-negative with a normalized
trace, one has that
\begin{equation}
\Vert r_1( \theta)\Vert_1 \, \leq \,
2 \left\| B_{a,T} ( \theta) - B_{a,T}^{(W)} ( \theta) \right\| \, \leq
\, C \, L^{2d} \, e^{- k \gamma_L L},
\end{equation}
using Lemma~\ref{lem:bbd} above. 

To estimate $r_2( \theta)$, we note that as in (\ref{eq:gsev}), 
\begin{equation} \label{eq:d1rho}
\partial_1 \rho_0( \theta, - \theta) \, = \, B_1 ( \theta) \, \rho_0( \theta, - \theta) \, + \, 
\rho_0( \theta, - \theta) \, B_1( \theta)^*,
\end{equation}
where $\partial_1$ denotes differentiation with respect to only the first
twist angle, namely $\theta$, which is situated near the sites $(m, V_L)$. 
Here we have also used that $ \partial_1E_0( \theta, - \theta) = 0$, 
see Lemma~\ref{lem:partialsE}. A simple norm estimate yields that
\begin{eqnarray} \label{eq:r2bd1}
\Vert r_2( \theta) \Vert_1 & \leq &
2 \left\| \, \left( \, B_{a,T}^{(W)} ( \theta) \, - \,  B_1 ( \theta)
      \right) P_0^{\theta} \, \right\|
\nonumber \\ & \leq & 2 \left\| \, B_{a,T} ( \theta) \, - \,
  B_{a,T}^{(W)} ( \theta) \, \right\| \, + \, 2 \left\| \, \left( \, B_{a,T} ( \theta) \, - \,  B_1 ( \theta)
    \right) P_0^{\theta} \, \right\|. 
\end{eqnarray}
Appealing again to Lemma~\ref{lem:bbd}, we see that $r_2( \theta)$ satisfies the
desired bound.  

Lastly, $ r_3( \theta) \, = \, {\rm Tr}_{m^c} \left[  \, \partial_2 \rho_0( \theta,
  -\theta) \right]$. Since we have shown in Lemma~\ref{lem:partialsE}
that  $\partial_2 E_0(\theta, - \theta) = 0$ as well, the analogue of 
(\ref{eq:d1rho}) holds for $\partial_2 \rho_0( \theta, -\theta)$. Thus,
\begin{eqnarray}
\Vert r_3( \theta) \Vert_1 & = &
\sup_{ \stackrel{O \in \A(m): }{\| O \| = 1}}  \left| {\rm Tr}
\left[ \, O \, \left( \, B_2 ( \theta) \, \rho_0(     \theta, - \theta) \, + \, 
\rho_0( \theta, -\theta) \, B_2( \theta)^* \, \right) \, \right]
\right| \nonumber \\ 
& \leq &  2 \, \sup_{ \stackrel{O \in \A(m): }{\| O \| = 1}} \int_0^{ \infty} \,
  \left| \, \langle \, \psi_0( \theta, - \theta), \, O \, \alpha_{it} \left( A_2(\theta)
  \right) \psi_0( \theta, - \theta)
  \,  \rangle \, \right| \, dt, 
\end{eqnarray}
where the observables $O$ are arbitrary elements of $\mathcal{A}(m)$,
again, the algebra of local observables with support in
$\Lambda_L^{(W)}(m)$. Integrals of this type are bounded using
Lemma~\ref{lem:itbd}; see also Remark~\ref{rem:intbd}. Since the 
observables we are considering have a separation distance proportional
to $L$, we may estimate
\begin{equation} \label{eq:r3bd}
\Vert r_3( \theta) \Vert_1 \, \leq \,
C \, L^{2d} \, e^{- k \gamma_L L}.
\end{equation}

Combining the results found on each of the remainders, 
we arrive at the estimate claimed in (\ref{thm:ptbd}).
\end{proof}

%
%
%
%

\subsection{Bound on the energy}

Equipped with Theorem~\ref{thm:statedif} and Lemma~\ref{lem:bbd}, we may now bound 
the excitation energy corresponding to the proposed state. 

\begin{thm} \label{thm:energyest}
Assume F1, F2, and LSM1-6.
If there exists a constant $c>0$ such that $\gamma_L L \geq c$ for
sufficiently large $L$ and we choose the parameters $a = \gamma_L / L$ and $T = L / 2$,
then there exists constants $C>0$ and $k>0$ so that
\begin{equation} \label{thm:ee}
\vert\langle \psi_1,  H_L  \psi_1 \rangle - E_0 \vert
\leq C L^{3d - 1} e^{- k \gamma_L L}\, 
\end{equation}
for large enough $L$. Here, we take $\psi_1 = \psi_{a,T}(2 \pi)$. 
\end{thm}

The proof of this theorem may be understood as follows. 
Recall that the ground state energy of the doubly twisted 
Hamiltonian is independent of $\theta$, i.e.,
\begin{equation}
E_0 \, = \, \langle \psi_0, H_L \psi_0 \rangle \, = \,  \langle
\psi_0(\theta, - \theta), H_{\theta, - \theta} \psi_0( \theta, - \theta) \rangle.
\end{equation}
Moreover, the separation between the twists of angle $\theta$ and $-\theta$
grows with the volume. Locality should enable one to estimate the energy
difference between performing two twists, the ground state, and
performing only one twist, the excited state. A rigorous version of
this idea is described below.

First, we recall some of the notation introduced in Section~\ref{sec:twham}. We
have written the twisted Hamiltonian as the sum of two terms 
\begin{equation} \label{eq:hamdecomp}
H_{\theta, -\theta} \, = \, H_{\theta, -\theta}^{(W)} \, + \, H^{(S)}.
\end{equation}
It is useful to further subdivide the twisted terms as
\begin{equation}
H_{\theta, -\theta}^{(W)} \, = \, H_{\theta}^{(W)}(m) \, + \, H_{-
  \theta}^{(W)} ( m + L/2),
\end{equation}
where $H_{\theta}^{(W)}(m)$ contains all those interaction terms in
$H_{\theta, -\theta}^{(W)}$ with support in a window about the twist
of angle $\theta$, i.e. $\Lambda_L^{(W)}(m)$, and
similarly, $H_{-\theta}^{(W)}(m+ L/2)$ contains all those interaction terms in
$H_{\theta, -\theta}^{(W)}$ with support in $\Lambda_L^{(W)} (m +
 L/2)$. The untwisted terms in (\ref{eq:hamdecomp}) are supported in
 the remaining strips. We refer to equations (\ref{eq:twham}) -(\ref{eq:hs}) for more
 details. It was also noted in Section~\ref{sec:twham} that  
\begin{equation}
W(\theta)^* \, H_{\theta, -\theta} \, W(\theta) \, = \, H.
\end{equation}
see (\ref{eq:unieq}).

Now, for any state $\psi$, one may calculate the expected energy due
to a single twist:
\begin{eqnarray}
\langle \psi, H_{\theta, 0} \psi \rangle & = &  \langle \psi, H_{\theta}^{(W)}(m) \psi \rangle
\, + \, \langle \psi, \left( H_0^{(W)} ( m + L/2)
\, + \, H^{(S)} \right) \psi \rangle  \nonumber \\ & = & 
E_0 \, + \, R_1( \theta) \, + \, R_2( \theta).
\end{eqnarray}
Here, we inserted appropriate terms so that we may compare $\langle \psi,
H_{\theta, 0} \psi \rangle$ to the ground state energy; the 
remainder terms are given by 
\begin{equation}
R_1( \theta) \,  := \, \langle \psi , H_{\theta}^{(W)}(m) \psi \rangle 
-  \langle \psi_0( \theta, - \theta ), H_{\theta}^{(W)}(m) \psi_0( \theta, - \theta) \rangle
\end{equation}
and
\begin{equation}
R_2( \theta) \,  := \, \langle \psi , \left( H_0^{(W)} ( m
    + L/2) \, + \, H^{(S)} \right) \psi \rangle 
-  \langle \psi_0,  \left( H_0^{(W)} ( m
    + L/2) \, + \, H^{(S)} \right) \psi_0 \rangle,
\end{equation}
where $\psi_0$ is the ground state of $H=H_{0,0}$. The
bound
\begin{equation} \label{eq:enerbd}
\left| \, \langle \psi, H_{\theta, 0} \psi \rangle \, - \, E_0 \, \right| \, \leq \, |R_1(
\theta)| \, + \, |R_2( \theta)|,
\end{equation}
readily follows for any state $\psi$. 

\begin{proof}(of Theorem~\ref{thm:energyest})
For each fixed $\theta$, the bound (\ref{eq:enerbd}) is valid for 
our proposed state $\psi_{a,T}(\theta)$. We will estimate the
resulting remainders uniformly for $\theta \in [0, 2 \pi]$ and thereby
prove the claimed result. 

To see this, we first rewrite the remainders in terms of the density matrices 
of the states restricted to the region containing the first twist.
It is clear that
\begin{eqnarray}
R_1( \theta) & = & \mbox{Tr} \left[ \, \left( \, \rho_{a,T}( \theta) - \rho_0( \theta, -
  \theta) \right) \, H_{\theta}^{(W)}(m) \, \right] \\  
 & = & \mbox{Tr}_m \left[ \, \mbox{Tr}_{m^c} \left[ \, \rho_{a,T}( \theta) - \rho_0( \theta, -
  \theta) \right] \, H_{\theta}^{(W)}(m) \, \right], \nonumber
\end{eqnarray}
where the partial traces are as defined just prior to Theorem~\ref{thm:statedif}. Thus 
\begin{eqnarray}
|R_1( \theta) | & \leq & \Vert H_{\theta}^{(W)}(m) \Vert \, \Vert \, \mbox{Tr}_{m^c} \left[ \rho_{a,T}(
      \theta) - \rho_0( \theta, - \theta) \right] \Vert_1 \\ 
      & \leq & C \, L^{3d-1} \, e^{- k \gamma_L L}, \nonumber 
\end{eqnarray} 
where we have used Theorem~\ref{thm:statedif}. In fact, from the
assumptions we have made, one verifies that
\begin{equation}
\| H_{\theta}^{(W)}(m) \| \, \leq \, 2 \sum_{x \in \Lambda_L^{(W)}(m)}
\sum_{X \ni x} \| \Phi(X) \| \, \leq \, 2 |||\Phi |||_1 \, |
\Lambda_L^{(W)}(m)| \, \leq \, C L^{d-1}.
\end{equation} 

For the second remainder,
\begin{equation} \label{eq:r2}
R_2( \theta) \, = \, \mbox{Tr} \left[ \, \left( \rho_{a, T}(
    \theta) - \rho_0( 0, 0 ) \right) \, \left( H_0^{(W)} ( m +
      L/2) \, + \, H^{(S)} \right) \, \right],
\end{equation}
we observe that the only $\theta$ dependence is in the density matrix
corresponding to the proposed state. Using the differential equation
(\ref{eq:exsev}), we find that 
\begin{eqnarray}
R'_2( \theta) & = & \mbox{Tr} \left(  \, \left[ B_{a,T} ( \theta), 
\rho_{a,T}( \theta) \right] \,  \left( H_0^{(W)} ( m +
      L/2) \, + \, H^{(S)} \right) \, \right)
\\ & = & - \,  \mbox{Tr} \left(  \, \left[ B_{a,T} ( \theta),  \left( H_0^{(W)}( m +
      L/2) \, + \, H^{(S)} \right) \right] \,
\rho_{a,T}( \theta) \, \right). \nonumber
\end{eqnarray}
The first term above is easy to estimate. Recall that the 
quantity $B_{a,T}^{(W)}( \theta)$ is supported in
  $\Lambda_L^{(W)}(m)$, whereas $H_0^{(W)}( m +
      L/2)$ has support in $\Lambda_L^{(W)} ( m +
      L/2)$. Thus       
\begin{equation}
\left[ B_{a,T}( \theta), H_0^{(W)}(m + L/2) \right] \, = \, 
\left[ B_{a,T}( \theta) \, - \, B_{a,T}^{(W)}( \theta), H_0^{(W)}(m + L/2) \right] ,
\end{equation}
and moreover,
\begin{eqnarray}
\left| \, \mbox{Tr} \left(  \, \left[ B_{a,T} ( \theta) \, - \,
      B_{a,T}^{(W)} ( \theta), H_0^{(W)}( m + L/2 ) \right] \, \rho_{a,T}( \theta) \, \right) \,
\right| \\  \leq \, 2 \, \left\| H_0^{(W)} ( m + L/2) \right\| \,   
\left\| \, B_{a,T}( \theta ) - B_{a,T}^{(W)}( \theta) \, \right\| \,
\leq \, C \, L^{3d-1} \, e^{-k \gamma L}. \nonumber
\end{eqnarray}
The second term may be similarly estimated. Let $\tilde{H}_{\theta,
  -\theta}^{(W)}$ be defined as in (\ref{eq:hamdecws}), excepting that the 
windows are slightly smaller: of size $\frac{L}{4} - 2R$. Then
$[\tilde{B}_{a,T}^{(W)}(\theta), H^{(S)}] = 0$, and the argument above applies. We have bounded $R_2(
\theta)$. \end{proof}

%
%
%
%
%

\section{Orthogonality} \label{sec:orthest}

We will now prove that, under the assumptions given in the introduction, the
proposed state is nearly orthogonal to the ground state.  As in Section
\ref{sec:ee}, we again make the assumption that $\gamma_L L\geq c>0$, for
sufficiently large $L$.

The reasoning behind orthogonality is simple. From LSM5, we know that
the ground state is an eigenvector of the translation
operator with eigenvalue 1, i.e., $T \psi_0 = \psi_0$. On the other
hand, the proposed state will very nearly be an eigenvector of $T_{2
  \pi, 0}$, as defined in Section~\ref{sec:twham}, with eigenvalue 1. 
Due to the odd parity condition $T_{2 \pi, 0} = -T$ and, hence, we find 
that the ground state and the proposed state are eigenvectors 
corresponding to distinct eigenvalues.

More concretely, it is easy to check that  
\begin{equation}
\langle \psi_{a, T}( 2 \pi), \psi_0 \rangle \, = \, \langle
T_{2 \pi, 0} \psi_{a, T}( 2 \pi), T \psi_0 \rangle + \langle
\left( I - T_{2 \pi,0} \right) \psi_{a, T}( 2 \pi), \psi_0 \rangle,
\end{equation}
from which the estimate
\begin{equation} \label{eq:orthy}
\left| \langle \psi_{a, T}( 2 \pi), \psi_0 \rangle \right| \,
\leq \, \frac{1}{2} \left\| \left( T_{2 \pi, 0} - I \right)
  \psi_{a, T}( 2 \pi) \right\|
\end{equation}
immediately follows. The remainder of this section will be used to prove a
bound on 
\begin{equation} \label{eq:orthoest}
\left\| \, T_{\theta, 0} \psi_{a,T}( \theta) - \psi_{a,T}(
   \theta) \, \right\| 
\end{equation} 
uniformly for $\theta \in [0, 2 \pi]$. This is the content
of Theorem~\ref{thm:orthoest}. 

%
%

\subsection{Observations concerning the twisted ground state}
We begin with a warm-up exercise involving the twisted ground state. 
In Section~\ref{sec:twham}, we saw that the twisted ground state 
is invariant with respect to the twisted translations; i.e., 
$T_{\theta, - \theta} \psi_0( \theta, - \theta) 
= \psi_0( \theta, - \theta)$, and therefore
\begin{equation} \label{eq:Tder=0}
\partial_{\theta} \left[ \, T_{\theta, - \theta} \psi_0( \theta, -
  \theta) \, - \, \psi_0( \theta, - \theta) \, \right] \, = \, 0.
\end{equation}
One may rewrite this derivative in the form of an operator acting on
$\psi_0( \theta, - \theta)$, i.e., (\ref{eq:Tder=0}) is equivalent to
\begin{equation} \label{eq:Der=0}
D(  \theta) \psi_0( \theta, - \theta) = 0,
\end{equation}
where $D(\theta)$ is given by
\begin{equation} \label{eq:defO}
D( \theta) \, = \, \partial_{\theta} T_{\theta, - \theta} \,
T_{\theta, -\theta}^* \, + \, T_{\theta, - \theta} \,
B(\theta ) \, T_{\theta, - \theta}^* \, - \,
B(\theta). 
\end{equation}
Here we have used the differential equation for $\psi_0( \theta, -
\theta)$, i.e. (\ref{eq:gstatede}), and the notation from the beginning
of Section~\ref{sec:locee}, which will be used throughout this section.

It will be easy to see that the operator $D(\theta)$
can be written as the sum of two terms, $D_1(\theta)$ and $D_2(\theta)$,
corresponding to the twists at $m$ and $m+L/2$, respectively. 
The goal of this subsection is to estimate 
$\Vert D_1(\theta)\psi_0(\theta, -\theta)\Vert$, see
Lemma~\ref{lem:d1bd} below.

Using (\ref{eq:twist}), one finds that 
\begin{equation}
\partial_{\theta} T_{\theta, - \theta} \,
T_{\theta, -\theta}^* \, = \, i \sum_{v \in V_L} S_{(m+1,v)}^3 \, - \,
 i \sum_{v \in V_L} S_{(m+L/2+1,v)}^3.
\end{equation}
One has that $D( \theta) = D_1( \theta) - D_2( \theta)$ where
\begin{equation}
D_1( \theta) \, = \,  i \sum_{v \in V_L} S_{(m+1,v)}^3 \, + \, T_{\theta, - \theta} \,
B_1( \theta) \, T_{\theta, - \theta}^* \, - \,
B_1(\theta), 
\end{equation}
and
\begin{equation}
D_2( \theta) \, = \,  i \sum_{v \in V_L} S_{(m+L/2+1,v)}^3 \, + \, T_{\theta, - \theta} \,
B_2( \theta) \, T_{\theta, - \theta}^* \, - \,
B_2(\theta), 
\end{equation}
For what follows, we will denote by $\langle A \rangle_{\theta} =
\langle \psi_0( \theta, - \theta), A \psi_0( \theta, - \theta)
\rangle$ the twisted ground state expectation of a local observable
$A$. We have demonstrated in Lemma~\ref{lem:partialsE} that 
\begin{equation}
0 \, = \, \partial_i E_0( \theta, - \theta)  \, =  
\, \langle \partial_i H_{\theta, -\theta} \rangle_{\theta} \, 
= \, \langle A_i(\theta) \rangle_{\theta},  
\end{equation} 
for $i = 1,2$. From this, we conclude that
\begin{equation}
\left\langle \, T_{\theta, -\theta} \, B_i(\theta) \, T_{\theta, -\theta}^* \,
\right\rangle_{\theta}  \, =  \, \left\langle \, B_i( \theta) \, \right\rangle_{\theta}  \, = \, \langle
A_i(\theta) \rangle_{\theta} \, = \, 0, 
\end{equation}
as well. Moreover, we similarly have that 
\begin{equation}
\langle D_i( \theta) \rangle_{\theta} \, = \, 0 \quad \quad \mbox{as}
\quad \quad \left\langle \sum_{v \in V_L} S_{(x,v)}^3
\right\rangle_{\theta} \, = \, \left\langle \sum_{v \in V_L} S_{(x,v)}^3
\right\rangle_0 \, = \, 0, 
\end{equation}
for any $x \in [1,L]$. For the last equality above, we used that $\psi_0( \theta, -\theta) = W(
\theta) \psi_0$, $W(\theta)$ commutes with the third component of the spins, rotation
invariance implies that the total spin is zero, and
translation invariance in the 1-direction. 

Since $D( \theta) \psi_0( \theta, -\theta) = 0$, we have that $D_1(
\theta) \psi_0( \theta, - \theta) = D_2( \theta) \psi_0( \theta, -
\theta)$ from which it is clear that
\begin{equation} \label{eq:d1=d2}
0 \, = \, \langle D(\theta)^*D( \theta) \rangle_{\theta} \, = \, 2 \langle D_1( \theta)^*
D_1(\theta) \rangle_{\theta} \, - \, 2 \langle D_1( \theta)^* D_2(
\theta) \rangle_{\theta}.
\end{equation}
As indicated above, we wish to estimate the first term on the right hand 
side above. We do so by estmating the second term. Observe that
\begin{eqnarray}
&&\langle D_1( \theta)^* D_2( \theta) \rangle_{\theta} \, = \, - \sum_{v, v'
  \in V_L} \langle S_{(m+1,v)}^3 \, S_{(m+L/2+1,v')}^3 \rangle_0 
\label{eq:d12exp} \\
&& +
\, i \sum_{v \in V_L} \int_0^{\infty} \left\langle \left(S_{(m,v)}^3
    \, - \, S_{(m+1,v)}^3 \right) \alpha_{it} \left( A_2(\theta)
  \right) \right\rangle_{\theta} \, dt  
\nonumber\\
&& + \, \int_0^{\infty} \int_0^{\infty}
\left\langle \alpha_{it}(A_1(\theta))^* \left(
    \alpha_{is}(A_2(\theta)) - \alpha_{is} \left( T_{\theta,
      -\theta}^* A_2(\theta)T_{\theta, -\theta} \right) \, \right)
\right\rangle_{\theta} \, ds \, dt 
\nonumber \\ 
&& + i\sum_{v' \in V_L} \int_0^{\infty} \left\langle
  \alpha_{it}(A_1(\theta))^* \left(S_{(m+ L/2+1,v')}^3
  - S_{(m+ L/2,v')}^3\right)
\right\rangle_{\theta} \, dt 
\nonumber\\
&& + \, \int_0^{\infty} \int_0^{\infty}
\left\langle \alpha_{it}(A_1(\theta))^* \left(
    \alpha_{is}(A_2(\theta)) - \alpha_{is} \left( T_{\theta,
      -\theta} A_2(\theta)T_{\theta, -\theta}^* \right) \, \right)
\right\rangle_{\theta} \, ds \, dt. \nonumber
\end{eqnarray}
That each of these terms is bounded follows from our decay of
correlations results found in Section~\ref{sec:gap}. In fact, we have proven the following lemma.
\begin{lem} \label{lem:d1bd} Assume F1, F2, and LSM1-6.
If there exists a constant $c>0$ such that $\gamma_L L \geq c$ for
sufficiently large $L$, then there exists constants $C>0$ and $k>0$ so
that
\begin{equation} \label{eq:d1bd}
\Vert D_1(\theta) \psi_0(\theta,-\theta)\Vert^2 \, \leq \, C L^{3d-1}
e^{-k \gamma_L L},
\end{equation}
for $L$ large enough.
\end{lem}
\begin{proof}
Clearly, one has that
\begin{equation}
\Vert D_1(\theta) \psi_0(\theta,-\theta)\Vert^2 
=\, \left\langle \, D_1( \theta)^* D_1(\theta) \,
  \right\rangle_{\theta} \, = \,  \left\langle \, D_1( \theta)^* D_2(\theta) \,
  \right\rangle_{\theta},
\end{equation}
from (\ref{eq:d1=d2}) above. Applying Theorem~\ref{thm:expclu}, Lemma~\ref{lem:itbd}, and 
Remark~\ref{rem:intbd}, as appropriate, to the terms found in (\ref{eq:d12exp}),
one arrives at (\ref{eq:d1bd}).
\end{proof}
%
%

\subsection{Orthogonality of the excited state}
We are now ready to provide the orthogonality estimate. 

\begin{thm} \label{thm:orthoest}
Assume F1, F2, and LSM1-6.
If there exists a constant $c>0$ such that $\gamma_L L \geq c$ for
sufficiently large $L$ and we choose the parameters $a = \gamma_L / L$ and $T = L / 2$,
then there exists constants $C>0$ and $k>0$ so that
\begin{equation} \label{thm:orest}
\left| \langle \psi_{a, T}( 2 \pi), \psi_0 \rangle \right| \,
\leq \,  C L^{2d} e^{- k \gamma_L L}
\end{equation}
when $L$ is large enough.
\end{thm}
\begin{proof} 
As is clear from (\ref{eq:orthy}), Theorem~\ref{thm:orthoest} follows
from bounding the quantity appearing in (\ref{eq:orthoest}) uniformly
for $\theta \in [0, 2 \pi]$. A short calculation, 
using (\ref{eq:exsde}), shows that 
\begin{equation} \label{eq:patevo}
\partial_{\theta} \left[ \, T_{\theta, 0} \psi_{a,T}( \theta) - \psi_{a,T}(
   \theta) \, \right] \, = \, C_{a, T}( \theta) \, \left[ \, T_{\theta, 0} \psi_{a,T}( \theta) - \psi_{a,T}(
   \theta) \, \right] \, + \, D_{a,T}(\theta) \, \psi_{a,T}( \theta), 
\end{equation}
where 
\begin{equation} \label{eq:catdef}
C_{a, T}( \theta) \, = \, \partial_{\theta} T_{\theta, 0}
\, T_{\theta, 0}^* \, + \,
T_{\theta, 0} \, B_{a, T}(\theta) \, T_{\theta,0}^*, 
\end{equation}
and
\begin{equation} \label{eq:datdef}
D_{a, T}( \theta) \, = \, \, \partial_{\theta} T_{\theta, 0}
\, T_{\theta, 0}^* \, + \,
T_{\theta, 0} \, B_{a, T}(\theta) \, T_{\theta,0}^*, \, - \,
B_{a, T}( \theta),  
\end{equation}
are both anti-Hermitian operators. The first term on the 
right hand side of (\ref{eq:patevo}) is norm-preserving,
and therefore, we need only bound the norm of the second by 
Theorem~\ref{thm:normp}. 

The norm of $D_{a,T}( \theta) \psi_{a,T}(\theta)$ will now be
estimated by rewriting it in terms of quantities for which 
we have already proven bounds. Each term will be shown 
to satisfy a bound of the form (\ref{thm:orest}). 

We begin by writing
\begin{eqnarray} \label{eq:normdat}
\| \, D_{a,T}( \theta) \, \psi_{a, T}( \theta) \, \|^2 &
= & \mbox{Tr} \left[ D_{a,T}( \theta)^* D_{a,T}( \theta) \,
  \rho_{a,T}( \theta) \right] \nonumber \\ &
= &   \mbox{Tr} \left[ D_{a,T}( \theta)^* D_{a,T}( \theta) \,
  \rho_0( \theta, -\theta) \right] 
\nonumber\\
&& + \,  \mbox{Tr} \left[ D_{a,T}( \theta)^* D_{a,T}( \theta) \,
  \left( \rho_{a,T}( \theta) \, - \, \rho_0( \theta, -\theta) \right) \right]. 
\end{eqnarray}
The first term on the right hand side above, which is equal to 
$\| D_{a,T}(\theta) \psi_0( \theta, -\theta) \|^2$, may be 
estimated by comparing it with the vector 
$D_1(\theta) \psi_0(\theta, -\theta)$ introduced in the previous
subsection. In fact, 
\begin{equation}
\| D_{a,T}( \theta) \psi_0( \theta, -\theta) \| \, \leq \,  \|
D_1( \theta) \psi_0( \theta, -\theta) \| \, + \, \| \left(D_{a,T}(
  \theta) \, - \, D_1( \theta) \right) \psi_0( \theta, -\theta) \|.
\end{equation}
We bounded the first term above in Lemma~\ref{lem:d1bd}. For the
second, observe that 
\begin{eqnarray}
D_{a,T}( \theta) \, - \, D_1( \theta) & = & T_{\theta, -\theta}
\left( B_{a,T}( \theta ) \, - \, B_1( \theta) \right) T_{\theta, -\theta}^*
\\  & \mbox{ } &  + \, T_{\theta, 0}
\left( B_{a,T}(\theta) \, - \, B_{a,T}^{(W)}( \theta) \right) T_{\theta, 0}^*
\nonumber \\  & \mbox{ } &  - \, T_{\theta, - \theta}
\left( B_{a,T}(\theta) \, - \, B_{a,T}^{(W)}( \theta) \right) T_{\theta,
  -\theta}^* \nonumber \\ & \mbox{ } & + \, B_1( \theta) \, - \, B_{a,T}(\theta), \nonumber
\end{eqnarray}
from which it is clear that
\begin{eqnarray}
\| \left(D_{a,T}( \theta) \, - \, D_1( \theta) \right) \psi_0(
\theta, -\theta) \| & \leq & 2 \, \| B_{a,T}(\theta) \, - \,  B_{a,T}^{(W)}(\theta)\| 
\nonumber\\ 
& \mbox{ } & + \, 2 \, \| \left(  B_{a,T}(\theta) \, - \,  B_1(\theta) \right) \, P_0^{\theta} \, \|. 
\end{eqnarray}
That each of these terms satisfies the desired bound 
follows from Lemma~\ref{lem:bbd}. 

For the final term on the right hand side of (\ref{eq:normdat}), we insert
and remove
\begin{equation}
D_{a, T}^{(W)}( \theta) \, = \, \, \partial_{\theta} T_{\theta, 0}
\, T_{\theta, 0}^* \, + \,
T_{\theta, 0} \, B_{a, T}^{(W)}(\theta) \, T_{\theta,0}^*, \, - \,
B_{a, T}^{(W)}( \theta), 
\end{equation}
a local observable supported in $\Lambda^{(W)}_L(m)$. Observe that
\begin{equation}
\| D_{a,T}^{(W)}( \theta) \| \, \leq \, \| \partial_{\theta}T_{\theta,0}
T_{\theta, 0}^* \| \, + \, 2 \| B_{a,T}^{(W)}(\theta) \| \, \leq \, C L^d,
\label{eq:dw}\end{equation}
where we have used Proposition \ref{prop:Bbound}.
We may write
\begin{eqnarray}
 \mbox{Tr} \left[ D_{a,T}( \theta)^* D_{a,T}( \theta) \,
  \left( \rho_{a,T}( \theta) \, - \, \rho_0( \theta, -\theta) \right)
\right] \, = \,   \mbox{Tr} \left[ D_{a,T}^{(W)}( \theta)^* D_{a,T}^{(W)}( \theta) \,
  \left( \rho_{a,T}( \theta) \, - \, \rho_0( \theta, -\theta) \right)
\right] \nonumber \\ \quad \quad + \,  \mbox{Tr} \left[ \left( D_{a,T}(
    \theta)^* D_{a,T}( \theta) \, -  D_{a,T}^{(W)}( \theta)^* D_{a,T}^{(W)}(
    \theta) \right) \,
  \left( \rho_{a,T}( \theta) \, - \, \rho_0( \theta, -\theta) \right) \right]
\end{eqnarray}

The first term above may be estimated by
\begin{eqnarray} 
&&\left|\mbox{Tr}_m \left[ \, D_{a,T}^{(W)}( \theta)^{*} \, D_{a,T}^{(W)}( \theta) \, 
  \mbox{Tr}_{m^c} \left[ \rho_{a,T}( \theta) \, - \, \rho_0( \theta, - \theta)
    \, \right] \, \right]\right|\\
&&\quad \leq 
\left\| D_{a,T}^{(W)}( \theta)^{*} \, D_{a,T}^{(W)}( \theta)
\right\| \, \left\Vert \mbox{Tr}_{m^c} \left[ \rho_{a,T}( \theta) \, 
- \, \rho_0( \theta, - \theta)
    \, \right] \right\Vert_1
\nonumber \\ 
&&\quad \leq C L^{4d} e^{-k\gamma L}\, , \nonumber
\end{eqnarray}
where for the final inequality above we used Theorem~\ref{thm:statedif}
again. 

For the second term, we rewrite the difference as
\begin{eqnarray}
&&D_{a,T}( \theta)^{*} \, D_{a,T}( \theta) -  
D_{a,T}^{(W)}( \theta)^{*} \, D_{a,T}^{(W)}( \theta) \,\nonumber\\
&&\quad =\, \left(  D_{a,T}( \theta) -  D_{a,T}^{(W)}( \theta) \right)^{
  *} \, D_{a,T}( \theta) \, 
+  D_{a,T}^{(W)}( \theta)^{*} \, \left(  D_{a,T}( \theta)
  -  D_{a,T}^{(W)}( \theta) \right),
\end{eqnarray}
and apply the norm estimate
\begin{equation} \label{eq:intdrop}
\left\| D_{a,T}( \theta) - D_{a,T}^{(W)}( \theta) \right\| \, \leq \, 2 \left\|
  B_{a,T}( \theta) - B_{a,T}^{(W)}(\theta) \right\|.
\end{equation}
We find that
\begin{eqnarray}
 \mbox{Tr} \left[ \left( D_{a,T}(
    \theta)^* D_{a,T}( \theta) \, -  D_{a,T}^{(W)}( \theta)^* D_{a,T}^{(W)}(
    \theta) \right) \,
  \left( \rho_{a,T}( \theta) \, - \, \rho_0( \theta, -\theta) \right)
\right] \, \leq \\ \nonumber 4 \left\|  B_{a,T}( \theta) -
  B_{a,T}^{(W)}(\theta) \right\| \,
\left( \| D_{a,T}( \theta) \| \, + \, \| D_{a,T}^{(W)}( \theta) \| \right),
\end{eqnarray}
which satisfies the required bound by Lemma~\ref{lem:bbd} and an estimate 
analogous to (\ref{eq:dw}). This completes the proof of
Theorem~\ref{thm:orthoest}. \end{proof}

%
%
%
%
%
%
%
%

\section{Auxiliary results} \label{sec:basest}

In this section, we collect a number of auxiliary results,
technical estimates as well as a few lemmas of a more general
nature, which are needed for the proofs in Sections \ref{sec:ee}
and \ref{sec:orthest}. 
 
We first recall the Lieb-Robinson bounds which are used to
demonstrate quasi-locality of the dynamics associated to general
quantum spin systems, see Theorem~\ref{thm:lr}. Then, we 
observe in Proposition~\ref{prop:difham} that these 
Lieb-Robinson bounds may be used to compare 
the dynamics of a Hamiltonian defined on a given system with 
the dynamics of the same Hamiltonian restricted to a 
subsystem. Next, we provide in Lemma~\ref{lem:lrbmt} an explicit
bound which applies to the specific type of interactions we consider
in this work.

In Section~\ref{subsec:itapp}, we introduce the operators 
$B_{a,T}(A, H)$ which play a prominent role in our argument. We first  
discuss a few of their basic properties, and then use 
Proposition~\ref{prop:difham} to estimate the difference that arises
in defining the operator with the full Hamiltonian as opposed to the
Hamiltonian restricted to a subsystems; this is the content of 
Lemma~\ref{lem:decup}. Lastly, we remark on exactly how this estimate
will be used in the main text. 

We review the Exponential Clustering Theorem in Section~\ref{sec:gap},
and use it to prove a technical estimate, see
Lemma~\ref{lem:itbd}. Moreover, in this section we also prove
Lemma~\ref{lem:gapbd}. This result provides an estimate on the
quantity $\Vert \left( B_{a,T}(A, H)-B(A,H) \right) P_0\Vert$ in terms of the 
 parameters $a$, $T$, and the spectral gap of $H$, see (\ref{eq:normdif}).
Here $P_0$ denotes the spectral projection onto the ground state of
$H$, and the bound is valid for local observables $A$ satisfying 
$P_0AP_0=0$.

Lastly, we formulate a statement concerning solutions of certain simple 
differential equations in Section~\ref{sec:npf}.   

%
%
%

\subsection{Lieb-Robinson bounds} \label{subsec:lrb} 
For what follows, we adopt the same general framework
for quantum spin models that was described in Section \ref{sec:setup},
including Conditions F1, F2,  and the assumption that 
$\Vert\Phi\Vert_\lambda < +\infty$ for some $\lambda >0$ (see (\ref{eq:defnphia})
for the definition of the norm $\Vert\cdot\Vert_\lambda$).

We will use the following version of the Lieb-Robinson bound
\cite{nachtergaele2006b}, which is a variant 
of the results proven in \cite{nachtergaele2006a,hastings2006}.

\begin{thm}[Lieb-Robinson Bound]\label{thm:lr}
Let $\lambda \geq 0$ and take $\Phi \in \mathcal{B}_{\lambda}( \Lambda)$. 
For any pair of local observables $A \in \mathcal{A}_{X}$ and 
$B \in \A_Y$ with $X, Y \subset \Lambda$, one may estimate  
\begin{equation} \label{eq:lrbd}
\left\| [ \alpha_t(A), B ] \right\| \, \leq \, \frac{2 \, \| A \|
\, \|B \|}{C_{\lambda}(F)} \, g_{\lambda}(t) \, \sum_{x \in X} \sum_{y \in
  Y} F_{\lambda} \left( d(x,y) \right),
\end{equation}
for any $t \in \mathbb{R}$. Here $\{ \alpha_t \}$ is the dynamics
generated by $\Phi$, and one may take
\begin{equation} \label{eq:gatt}
g_{\lambda}(t) \, = \, \left\{ \begin{array}{cc} 
\left(e^{2 \, \| \Phi \|_{\lambda} \, C_{\lambda}(F) \, |t|} - 1 \right)  & \mbox{ if }
d(X,Y)>0, \\  e^{2 \, \| \Phi \|_{\lambda} \, C_{\lambda}(F) \, |t|} & \mbox{
  otherwise.} \end{array} \right.
\end{equation}
\end{thm}

Our proof of the Lieb-Schultz-Mattis theorem relies heavily on comparing the
time evolution corresponding to a given Hamiltonian to that of the
Hamiltonian restricted to a subsystem. The errors that result
from such a comparison can be estimated in terms of a specific
commutator to which the Lieb-Robinson bounds readily apply.

We begin with some notation. Let $\lambda \geq 0$ and consider $\Phi
\in \mathcal{B}( \Lambda)$. For finite $\Lambda_0 \subset \Lambda$,
the Hamiltonian corresponding to $\Phi$ restricted to $\Lambda_0$ is
given by the self-adjoint operator
\begin{equation}
H_0 \, = \, \sum_{X \subset \Lambda_0} \Phi(X).
\end{equation}
We will denote by $\alpha_t^{(0)}$ the time evolution corresponding to
$H_0$, i.e., for any local observable $A$, 
$\alpha_t^{(0)}(A) = e^{itH_0}Ae^{-itH_0}$ for all $t \in \mathbb{R}$.

\begin{prop} \label{prop:difham} Let $\lambda \geq 0$ and $\Phi \in
  \mathcal{B}_{\lambda}( \Lambda)$. Suppose the Hamiltonian
  corresponding to $\Phi$ restricted to a finite volume $\Lambda_0
  \subset \Lambda$ is written as the sum of two self-adjoint
  operators, i.e., $H_0 = H_1 + H_2$. Denoting by $\alpha_t^{(i)}$ the
  time evolution corresponding to $H_i$, for $i=0,1,2,$ then for any 
  local observable $A$ and $t \in \mathbb{R}$, one has that
\begin{equation} \label{eq:difdyn}
\| \, \alpha_t^{(0)}(A) \, - \, \alpha_t^{(1)}(A) \,  \| \, \leq
\, \int_0^{|t|} \, \left\| [ H_2 , \alpha_s^{(1)}(A) ] \right\|
\, ds.
\end{equation}
\end{prop}

\begin{proof}
Define the function $f: \mathbb{R} \to \mathcal{A}$ by
\begin{equation}
f(t) \, := \, \alpha_t^{(0)}(A) \, - \, \alpha_t^{(1)}(A).
\end{equation}
A simple calculation shows that $f$ satisfies the following differential equation:
\begin{equation}
f'(t) = \, i
\left[ H_0 - H_1, \alpha_t^{(1)}(A) \right] \, + \, i  \left[ H_0, f(t) \right],
\end{equation}
subject to the boundary condition $f(0)=0$. As this is a first order
equation, the solution can be found explicitly:
\begin{equation} \label{eq:desol}
f(t) \, = \, \alpha_t^{(0)} \left( \, \int_0^t \, \alpha_{-s}^{(0)} \left(
  \, i \, \left[ H_2, \alpha_s^{(1)}(A) \right] \, \right) \, ds \, \right).
\end{equation}
Using expression (\ref{eq:desol}) and the automorphism property of $\alpha^{(0)}_t$, it is clear that
\begin{equation}\label{normft}
\| \, f(t) \,  \| \, \leq
\, \int_0^{|t|} \, \left\| [ H_2, \alpha_s^{(1)}(A) ] \right\|
\, ds,
\end{equation}
as claimed.
\end{proof}

To estimate the norm of the commutator appearing in Proposition~\ref{prop:difham},
specifically in the bound (\ref{eq:difdyn}), it is useful to specialize 
the general Lieb-Robinson bounds described above to 
the exact context we encounter in the present work.
For example, we will be interested in specific finite volume 
Hamiltonians, those defined in Section~\ref{sec:construction} as $H_{\theta, \theta'}$, and
particular observables, such as $A_1(\theta) = \partial_{1} H_{\theta, -\theta}$ 
and $A_2(\theta) = \partial_{2} H_{\theta, - \theta}$. 
Let $\alpha_t$ be the time evolution corresponding to the $H_{\theta,
\theta'}$ Hamiltonian, and let $\alpha_t^{(W)}$ denote the 
dynamics associated with the Hamiltonian $H_{\theta, \theta'}^{(W)}$
which is defined in (\ref{eq:hamdecws}). We use the following estimate several
times.

\begin{lem}\label{lem:lrbmt} Let $\Phi \in \mathcal{B}_{\lambda}(
  \Lambda)$, then there exists constants $C>0$ and $k>0$ for which
\begin{equation} \label{eq:lrbmt}
\max_{i = 1,2} \sup_{\theta \in [0, 2\pi]} \left\| \, \left[ H^{(S)},
      \alpha_t^{(W)} \left( A_i (\theta) \right) \right]
  \, \right\| \, \leq \, C  \, e^{k |t|} \, L^{2(d-1)} \, e^{- \lambda L/4}.
\end{equation}
Here it is  important that $C$ and $k$ depend only on the
properties of the underlying set $\Lambda$ and the interaction $\Phi$;
they do not depend on the length scale $L$.
\end{lem}
\begin{proof}
We will estimate the above commutator in the case that the observable
is $A_1(\theta)$; an analogous result holds for $A_2(\theta)$. Recall that in (\ref{eq:hs})
we wrote $H^{(S)}$ as a sum of interaction terms. Similarly, if
one denotes by $P_m( \theta; Y) := V_m( \theta)^* \Phi(Y) V_m(
\theta) - \Phi(Y)$, then $A_1(\theta)$ may be written as 
\begin{equation}
A_1(\theta) \, = \, \sum_{Y \subset \Lambda_L}
\partial_{\theta} P_m( \theta; Y) \, = \, - i \, \sum_{\stackrel{Y \subset \Lambda_L:}{P_m( \theta; Y) \neq 0}}
\sum_{y \in Y_+} V_m( \theta)^* \left[ S_y^3, \Phi(Y) \right] V_m( \theta)\, ,
\end{equation}
where $Y_+$ is the set of sites $y\in Y$ strictly to the right of $m$.
Inserting both of these expressions into the right hand side of
(\ref{eq:lrbmt}) and applying the triangle inequality, it is clear
that we must bound many terms of the form
\begin{equation} \label{eq:baseterm}
\left\| \, \left[ \,\Phi(X), \, \alpha_t^{(W)} \left( V_m(\theta)^*
      \left[ \, S_y^3, \, \Phi(Y) \, \right] \, V_m( \theta) \,
    \right) \, \right] \, \right\|. 
\end{equation}
Term by term, we apply the Lieb-Robinson bound provided by
Theorem~\ref{thm:lr}, and use that the distance between the 
supports of $X$ and $Y$ is linear in $L$; concretely for any $x \in X$ and $y
\in Y$, $d(x,y) \geq d(X,Y) \geq \frac{L}{4} - 3R$. We find that each
term described by (\ref{eq:baseterm}) satisfies an upper bound of the
form
\begin{equation}
C(t) \, \| \Phi(X) \| \, |Y| \, \| [S_y^3, \Phi(Y)] \| \, e^{- \lambda L/4},
\end{equation} 
where $C(t)$ may be taken as
\begin{equation}
C(t) = \frac{2 \|F\|}{C_{\lambda}(F)} e^{2 C_{\lambda}(F) \| \Phi
  \|_{\lambda} |t| + 3 \lambda R}.
\end{equation}
We need only count the number of terms. The combinatorics of the sums
may be naively estimated as follows: $H^{(S)}$ corresponds to a sum of
the form
\begin{equation}
\sum_{\stackrel{X \subset \Lambda_L:}{X \cap \Lambda_L^{(S)}
  \neq \emptyset}} \, \leq \, \sum_{n = \frac{L}{4} - R + 1}^{\frac{L}{4} + R - 1}
\sum_{v \in V_L} \sum_{X \ni (m+n,v)} \, + \,  \sum_{n = \frac{3L}{4} - R + 1}^{\frac{3L}{4} + R - 1}
\sum_{v \in V_L} \sum_{X \ni (m +n,v)},
\end{equation}
whereas for $A_1( \theta)$ we have that the sum
\begin{equation}
\sum_{ \stackrel{Y \subset \Lambda_L:}{P_m( \theta; Y) \neq 0}}
\sum_{y \in Y} \, \leq \, \sum_{n=m-R}^{m+R} \, \sum_{v \in V_L}
\sum_{Y \ni (n, v)} \sum_{y \in Y}. 
\end{equation}
Putting everything together, we have obtained that
\begin{equation}
\left\| \left[ H^{(S)}, \alpha_t^{(W)} \left( A_1 (\theta) \right) \right] \right\| \, \leq
\, 2 \, C(t) \, ||| \Phi |||_1 \, ||| \Phi |||_2 \, |V_L|^2 \, (2R+1)(2R-1) \, e^{- \lambda \frac{L}{4}},
\end{equation}
which proves the claim. Recall,
\begin{equation}
||| \Phi |||_1 := \sup_{x \in \Lambda} \sum_{X \ni x} \| \Phi(X) \|
\end{equation}
and
\begin{equation}
||| \Phi |||_2 := \sup_{x \in \Lambda} \sum_{X \ni x} |X| \sum_{x' \in X}
 \| \, [S_{x'}^3, \Phi(X)] \, \|.
\end{equation}
\end{proof}

%
%
%
%

\subsection{Approximation of the imaginary time evolution} \label{subsec:itapp}
For our proof of the Lieb-Schultz-Mattis Theorem, we introduce an
operator which, under certain assumptions, approximates  
the imaginary time evolution corresponding to a given Hamiltonian. 
In this section, we provide several basic estimates of this operator
to which we will often refer in the main text. 

Let $\lambda \geq 0$, $\Phi \in \mathcal{B}_{\lambda}( \Lambda)$,
$\Lambda_0 \subset \Lambda$ be a finite set, and $H$ be the
Hamiltonian corresponding to $\Phi$ restricted to $\Lambda_0$. 
Denote by $\alpha_t$, for $t \in \mathbb{R}$, 
the time evolution determined by $H$. For any local
observable $A$, $a >0$, $M>0$, and $t \neq 0$, define
\begin{equation} \label{eq:defAM}
A_{ a, M}(it, H) \, =  \, \frac{ e^{ - a t^2} }{2 \pi i} \,
\int_{-M}^{M} \, \alpha_s(A) \, \frac{e^{ - a s^2}}{s -
    i t} \, ds,
\end{equation}
and set  $A_{a}(it, H) = \lim_{M \to \infty}A_{a,M}(it,H)$.
We use the operator
\begin{equation} \label{eq:defbat0}
B_{a,T}(A, H) \, = \, - \, \int_0^T A_{a}(it, H) -
A_{a}(it, H)^* \, dt,
\end{equation}
to define our variational state in the main text, see (\ref{eq:exsde}).
We begin with some basic properties.

\begin{prop}[Shanti's Bound]\label{prop:Bbound} 
Let $\Phi \in \mathcal{B}_{\lambda}(\Lambda)$, $A$ be a local
observable, $a>0$, and $T>0$. The operator $B_{a,T}(A, H)$ is
  anti-hermitian and bounded. In particular, 
\begin{equation} \label{eq:batbd}
\| B_{a,T}( A, H) \| \, \leq \, \frac{ \| A \|}{2} \, \sqrt{ \frac{\pi}{a}}.
\end{equation}
\end{prop}

It is important to note that the bound above is independent of the
finte volume $\Lambda_0$ on which the Hamiltonian $H$ is defined.

\begin{proof} That $B_{a,T}(A, H)$ is anti-hermitian follows
  immediately from (\ref{eq:defbat0}). Combining (\ref{eq:defAM}) and
  (\ref{eq:defbat0}), one finds that
\begin{equation}
B_{a,T}(A, H) \, = \, \frac{i}{ \pi} \int_0^T
\int_{-\infty}^{\infty} e^{-a(s^2+t^2)} \, \alpha_s(A) \,
\frac{s}{s^2+t^2} \, ds \, dt,
\end{equation}
from which (\ref{eq:batbd}) easily follows as
\begin{equation}
\| B_{a,t}(A, H) \| \leq \frac{ \| A \|}{ \pi}
\int_{-\infty}^{\infty} e^{-as^2} \, |s| \int_0^T \frac{1}{s^2 + t^2} \,
dt \, ds \, \leq \, \frac{\|A \|}{2} \, \sqrt{ \frac{\pi}{a}}.
\end{equation}
\end{proof}
In situations where the local observable $A$ and the Hamiltonian $H$
are fixed, we will often write $A_{a}(it)$ and $B_{a,T}$ to
simplify notation. The following estimate is a simple consequence of 
(\ref{eq:defAM}). 
\begin{prop} \label{prop:ftime} Let $\Phi \in \mathcal{B}_{\lambda}( \Lambda)$
  and $A$ be a local observable. One has that
\begin{equation} \label{eq:ftbd}
\left\| \, \int_0^T \, A_{a}(it) - A_{a, M}(it) \, dt \,
\right\| \, \leq \, \frac{T}{2M} \, \frac{ \| A \|}{ \sqrt{ \pi
    a}} \, e^{ - a M^2}.
\end{equation}
\end{prop}

\begin{proof} 
For any $t \neq 0$,
\begin{equation}
A_{a}(it) - A_{a, M}(it) =   \frac{e^{- a
    t^2}}{ 2 \pi i} \, \int_{|s|>M} \alpha_s(A) \, \frac{e^{ -
  a s^2}}{s-it} \, ds,
\end{equation}
and therefore, one has the pointwise estimate
\begin{equation}
\left\| \, A_{a}(it) - A_{a, M}(it) \, \right\|  \, \leq \,
e^{- a t^2} \, \frac{ \| A \| }{ 2 \pi M} \, e^{ - a M^2} \,
\sqrt{ \frac{ \pi}{ a}}.
\end{equation}
Upon integration, (\ref{eq:ftbd}) readily follows. 
\end{proof}

We will now prove an analogue of Proposition~\ref{prop:difham} for the
quantities $B_{a,T}(A, H)$ introduced in (\ref{eq:defbat0}). The estimate
provided below is made explicit in terms of an a priori input, an assumed form
of the Lieb-Robinson bound, see (\ref{eq:lrest}) below.

\begin{lem} \label{lem:decup} Let $\lambda \geq 0$ and $\Phi \in
  \mathcal{B}_{\lambda}( \Lambda)$. Suppose the Hamiltonian
  corresponding to $\Phi$ restricted to a finite volume $\Lambda_0
  \subset \Lambda$ is written as the sum of two self-adjoint
  operators, i.e., $H_0 = H_1 + H_2$. Denote by $\alpha_t^{(i)}$ the
  time evolution corresponding to $H_i$, for $i=0,1,2.$ 
  If, for a given local observable $A$, there exists 
  numbers $c_i >0$, $i =1,2,3$, for which 
\begin{equation} \label{eq:lrest}
\left\|  \, [ \, H_2, \, \alpha^{(1)}_t(A) \, ] \, \right\| \, \leq \,
c_1 \, e^{ c_2 |t| \, - \, c_3},  
\end{equation}
for all $t \in \mathbb{R}$, then the following estimate holds 
\begin{equation} \label{lem:decupest}
\left\| \, B_{a,T}(A, H_0) - B_{a,T}(A, H_1) \,
\right\| \, \leq \, \frac{2T}{M} \, e^{- a M^2} \left( \, \frac{ \| A \|}{ \sqrt{ \pi
    a}} \, + \, \frac{c_1 M^2}{ \pi} \, \right),
\end{equation}
where $M$ has be chosen as the positive solution of 
\begin{equation} \label{eq:quadeq}
a M^2 + c_2 M -c_3=0.
\end{equation}
\end{lem}

We note that in our applications the numbers $c_i$ will 
depend on the observables $A$ and $H_2$;
in fact, they will be functions of the length scale $L$. We articulate
this dependence explicitly in Remark~\ref{rem:decup} below.

\begin{proof}
One may write
\begin{eqnarray}
B_{a,T}(A, H_0) - B_{a,T}(A, H_1)  & = & 
- \, \int_0^T \, A_{a}(it, H_0) - A_{ a}(it, H_1) \, dt \, \\
\mbox{ } & + & \int_0^T \, A_{a}(it, H_0)^* - A_{
  a}(it, H_1)^* \, dt, \nonumber 
\end{eqnarray}
and therefore
\begin{equation}
\left\| \, B_{a,T}(A, H_0) - B_{a,T}(A, H_1) \, \right\| \, \leq
\, 2 \, \left\|  \, \int_0^T \, A_{a}(it, H_0) - A_{
    a}(it, H_1) \, dt \, \right\|.  
\end{equation}
Moreover, the integrand may be expressed as 
\begin{eqnarray}
A_{a}(it, H_0) - A_{ a}(it, H_1)  &  = & 
A_{a}(it, H_0) - A_{ a, M}(it, H_0) + \nonumber \\
&  \mbox{ } & A_{a,M}(it, H_0) - A_{ a, M}(it, H_1) + A_{ a,
  M}(it, H_1) - A_{ a}(it, H_1), 
\end{eqnarray}
and for $j=0,1$, the bounds 
\begin{equation} \label{eq:propbd}
\left\| \, \int_0^T \, A_{a}(it, H_j) - A_{a, M}(it, H_j) \, dt \,
\right\| \, \leq \, \frac{T}{2M} \, \frac{ \| A \|}{ \sqrt{ \pi
    a}} \, e^{ - a M^2},
\end{equation}
follow immediately from Proposition~\ref{prop:ftime}. From this we
conclude that for any $M>0$,
\begin{eqnarray}
\left\| \, B_{a,T}(A, H_0) - B_{a,T}(A, H_1) \, \right\| \, 
&\leq&
\, 2 \, \left\|  \, \int_0^T \, A_{a,M}(it, H_0) - A_{
    a,M}(it, H_1) \, dt \, \right\|\nonumber\\ 
&&+ \frac{2T}{M} \, \frac{ \| A \|}{ \sqrt{ \pi
    a}} \, e^{ - a M^2} \, .  
\end{eqnarray}

Clearly, the pointwise estimate 
\begin{equation}
\| A_{a,M}(it, H_0) - A_{ a, M}(it, H_1) \| \leq
\frac{e^{- a t^2}}{2 \pi} \int_{-M}^M \, \frac{\| \alpha_s^{(0)}(A) -
\alpha_s^{(1)}(A) \|}{|s|} \, e^{- a s^2} \, ds,
\end{equation}
follows directly from (\ref{eq:defAM}). By Proposition~\ref{prop:difham}, we have that
\begin{equation}
 \| \alpha_s^{(0)}(A) - \alpha_s^{(1)}(A) \| \leq
\, \int_0^{|s|} \, \left\| [ H_2 , \alpha_x^{(1)}(A) ] \right\|
\, dx,
\end{equation}
and by assumption (\ref{eq:lrest}), the integrand satisfies a uniform
bound for $|s| \leq M$. The implication is that for all $|s| \leq
M$,
\begin{equation}
\frac{\| \alpha_s^{(0)}(A) - \alpha_s^{(1)}(A) \|}{|s|} \, \leq \, c_1
e^{c_2M -c_3}. 
\end{equation}
Putting everything together, we obtain that
\begin{eqnarray} 
\left\| \, B_{a,T}(A, H_0) - B_{a,T}(A, H_1) \,
\right\| & \leq & \frac{2T}{M} \, \frac{ \| A \|}{ \sqrt{ \pi
    a}} \, e^{ - a M^2} \nonumber \\ & + & \frac{c_1e^{c_2M-c_3}}{\pi}
\int_0^T \int_{-M}^M \, e^{ - a (t^2 +s^2)} \, ds dt. 
\end{eqnarray}
As $M$ here was arbitrary, we choose it as the (positive) solution
of the following quadratic equation $a M^2 + c_2 M -c_3=0$. In this case,
\begin{equation}
\left\| \, B_{a,T}(A, H_0) - B_{a,T}(A, H_1) \,
\right\| \, \leq \, \frac{2T}{M} \, e^{- a M^2} \left( \, \frac{ \| A \|}{ \sqrt{ \pi
    a}} \, + \, \frac{c_1 M^2}{ \pi} \, \right)
\end{equation}
as claimed.
\end{proof}

\begin{Remark} \label{rem:decup} In the main text of the paper, we will use
  Lemma~\ref{lem:decup} for the Hamiltonians $H_0 =
  H_{\theta, -\theta}$ and $H_1 = H_{\theta, -\theta}^{(W)}$ each of
  which depends on the length scale $L$;
  see Section~\ref{sec:construction} for the 
  relevant definitions. It is assumed that 
  $H_0$ has a gap $\gamma_L >0$ above the ground state energy. 
  The local observable $A$ will be exactly as in 
  Lemma~\ref{lem:lrbmt}, and therefore, the numbers $c_i$, $i=1,2,3$,
  may be taken as follows: $c_1 = C L^{2(d-1)}$, $c_2 = k$, and 
  $c_3 = \lambda L/4$, where again $C$ and $k$ depend
  only on the interaction and the underlying set $\Lambda$. In this case, we will choose the parametrization 
  $a = \gamma_L / L$ and $T = L /2 $. With this choice, the estimate 
  (\ref{lem:decupest}) takes the form:
\begin{equation}
\sup_{\theta \in [0, 2\pi]} \left\| \, B_{a,T} (\theta) - B_{a,T}^{(W)}(\theta) \, 
\right\|   \leq \, C \, L^{2d} e^{- k \gamma_L L} \left( 1 + 
  \frac{1}{ L^{d} \sqrt{ \gamma_L L}} \right).
\end{equation}  
Here we have used the notation from Section~\ref{sec:ee} and the fact 
that the gap $\gamma_L$ has a uniform bound from
above; see (\ref{eq:ugb}) in Section~\ref{sec:intro}.
\end{Remark}


\subsection{Estimates for gapped systems}\label{sec:gap}

We derive two useful results in this subsection. For the first we recall the
Exponential Clustering Theorem \cite{nachtergaele2006a,hastings2006}, and
use it to prove a technical estimate Lemma \ref{lem:itbd}.

The second crucial estimate in this subsection, Lemma \ref{lem:gapbd} below,
applies specifically to gapped systems. It provides a bound on the
norm of the difference in the operators $B_{a,T}(A,H)$ and $B(A,H)$
when restricted to the space of ground states corresponding to
$H$. The bound applies to local observables $A$ which project off the
ground state, i.e. satisfy $P_0AP_0 =0$ where $P_0$ is the spectral 
projection of $H$ onto the ground states, and is explicit in the 
parameters $a$, $T$, and the spectral gap of $H$, see (\ref{eq:normdif}) below.  

We will consider Hamiltonians $H$, of the type introduced in
section ~\ref{subsec:lrb}, with an additional feautre: a gap above the
ground state energy. To state the gap condition precisely, we consider
a representation of the system on a Hilbert space $\mathcal{H}$. 
This means that there
is a representation $\pi:\mathcal{A}\to\mathcal{B}(\mathcal{H})$, and a self-adjoint operator $H$ on
$\mathcal{H}$ such that
$$
\pi(\alpha_t(A))=e^{itH}\pi(A)e^{-it H}, 
$$
for all $t \in \mathbb{R}$ and $A \in \mathcal{A}$. For the results which follow, we will assume that 
$H \geq 0$ and that $\Omega \in \mathcal{H}$ is a normalized ground
state, i.e., a vector state for which $H \Omega =0$ and $\| \Omega \|
=1$.  We say that the system has a spectral gap in this 
representation if there exists $\delta >0$ such that $\sigma (H) \cap
(0,\delta) =\emptyset$, where $\sigma(H)$ is the spectrum of the
operator $H$. In that case, the spectral gap, $\gamma$, is defined
to be 
\begin{equation}
\gamma=\sup\{\delta > 0 \mid \sigma(H) \cap (0,\delta) =\emptyset\}.
\end{equation}
Let $P_0$ denote the orthogonal projection onto $\ker H$. From now on,
we will work in this representation and simply write $A$ instead of
$\pi(A)$. 

The following result concerning exponential clustering was proven in
\cite{nachtergaele2006a}.
\begin{thm}[Exponential Clustering] \label{thm:expclu}
Fix $\lambda > 0$. Let $\Phi \in \mathcal{B}_{\lambda}(\Lambda)$ be an
interaction for which the corresponding self-adjoint Hamiltonian 
has a representation $H \geq 0$ with a normalized ground 
state vector $\Omega$, i.e., $H \Omega = 0$ and $ \| \Omega \| = 1$. 
Let $P_E$ denote the family of spectral projections corresponding to
$H$. If $H$ has a spectral gap of size $\gamma >0$ above the 
ground state energy, then there exist $\mu >0$
such that for any local observables $A$ and $B$ with $A \in
\mathcal{A}_X$, $B \in \mathcal{A}_Y$, $d := \mbox{dist}(X,Y)>0$, and 
$P_0 B \Omega = P_0 B^* \Omega = 0$, the estimate 
\begin{equation}
\left| \, \left\langle \, \Omega, \, A \alpha_{it}(B) \Omega \,
  \right\rangle \, \right| \, \leq \, C(A,B) \, e^{- \mu d \, \left( \,
    1 \, + \, \frac{ \gamma^2 t^2}{4 \mu^2 d^2} \, \right)},
\end{equation}
holds for all $t \ : \, 0 \, \leq \, t (4 \| \Phi \|_{\lambda} C_{\lambda} + \gamma ) \, \leq \,
2 \, \lambda \, d$. Here, one may choose
\begin{equation}
C(A,B) = \| A \| \, \| B \| \, \left( \, 1 \, + \, \frac{2 }{ \pi \,
    C_{\lambda}} \, \sum_{x \in X} \sum_{y \in Y} F \left( d(x,y) \right) + \, \frac{1}{
    \sqrt{ \pi \mu d}} \, \right) 
\end{equation}
and
\begin{equation}
\mu = \frac{ \lambda \, \gamma }{ 4 \| \Phi \|_{\lambda} \, C_{\lambda} \, + \, \gamma}.
\end{equation}
\end{thm}

The above result easily leads to estimates on 
integrals of these ground state expectations. 
We state two such bounds in the next lemma, as they will arise in the
proof of our main result.

\begin{lem} \label{lem:itbd} Under the assumptions of
  Theorem~\ref{thm:expclu}, we have the estimates
\begin{equation} \label{eq:itbd1}
 \int_0^{\infty} \, \left| \, \langle \Omega, A \alpha_{it}(B) \Omega
  \rangle \, \right| \, dt   \, \leq \,  \left( \, 2 \mu d \, C(A,B) \, + \, \| A \| \, \|
    B \| \,  e^{- \mu \, d} \, \right) \, \frac{e^{- \mu d }}{\gamma}. 
\end{equation}
and
\begin{equation} \label{eq:itbd2}
 \int_0^{\infty} \, \int_0^{ \infty} \, \left| \, \langle \Omega, A \alpha_{i(s+t)}(B) \Omega
  \rangle \, \right| \, ds \, dt   \, \leq \,  \left[ \, \left( \mu d \right)^2 \, C(A,B) \, + \, \| A \| \, \|
    B \| \,  \left( 2 \mu d \, + \, e^{- \mu \, d} \right) \, \right] \, \frac{e^{- \mu d }}{\gamma^2}. 
\end{equation}
\end{lem}
\begin{proof}

Define $T$ by the equation $\gamma \, T = 2 \, \mu \, d$. 
We have that
\begin{equation}
 \int_0^T \, \left| \, \langle \Omega, A \alpha_{it}(B) \Omega
  \rangle \, \right| \, dt   \, \leq \, C(A,B) \, T \, e^{- \mu \, d},
\end{equation}
and also 
\begin{equation}
 \int_T^{\infty} \, \left| \, \langle \Omega, A \alpha_{it}(B) \Omega
  \rangle \, \right| \, dt  \, \leq \, \frac{\|A \| \, \| B \|}{ \gamma}
\, e^{- \gamma \, T}. 
\end{equation}
Combining these two bounds, we arrive at (\ref{eq:itbd1}). 
Similarly, one may estimate
\begin{equation}
\int_0^{T/2}\int_0^{T/2} \, \left| \, \langle \Omega, A \alpha_{i(s+t)}(B) \Omega
  \rangle \, \right| ds \, dt \, \leq \, \frac{C(A,B) \, T^2}{4} \, e^{- \mu \, d},
\end{equation}
\begin{equation}
 \int_{T/2}^{\infty}\int_0^{T/2} \, \left| \, \langle \Omega, A \alpha_{i(s+t)}(B) \Omega
  \rangle \, \right| \, ds \, dt  \, \leq \, \frac{\| A \| \, \|B \|
  \, T}{2 \gamma} \, e^{- \mu \, d},
\end{equation}
and finally,
\begin{equation}
\int_{T/2}^{\infty} \int_{T/2}^{\infty} \, \left| \, \langle \Omega, A \alpha_{i(s+t)}(B) \Omega
  \rangle \, \right| \, ds \, dt \, \leq \, \frac{\|A \| \, \| B \|}{ \gamma^2}
\, e^{- \gamma \, T}. 
\end{equation}
\end{proof}

\begin{Remark} \label{rem:intbd} In our applications, the Hamiltonian
  $H = H_{\theta, -\theta}$ depends on a length scale $L$ and has a
  gap $\gamma_L>0$ above the ground state energy. The support of the 
  observables $A$ and $B$ will have a minimal distance $d= L/2 -2R-1$,
  and moreover, $B=B(\theta)$ will either be $A_1( \theta)$ or
  $A_2(\theta)$. In this case, $\langle B(\theta) \rangle_{\theta} =
  0$ by Lemma~\ref{lem:partialsE}, and therefore, the assumptions of 
  Theorem~\ref{thm:expclu} hold. Here we have used $\langle \cdot
  \rangle_{\theta}$ to denote the ground state expectation
  corresponding to $\psi_0( \theta, -\theta)$ It is easy to see that 
  there exist positive constants $C'$ and $C''$ for which $C'
  \gamma_L L \leq \mu d \leq C'' L$, and thus ultimately constants $C$ and
  $k$ for which the bounds appearing in (\ref{eq:itbd1}) and 
  (\ref{eq:itbd2}) may be estimated by
\begin{equation} \label{eq:itbd12}
 \sup_{\theta \in [0, 2\pi]} \int_0^{\infty} \, \left| \, \langle A
   \alpha_{it} \left( B (\theta) \right) \rangle_{\theta} \, \right| \, dt   \, \leq \, C \, \| A \| \, |X| \,
\frac{L^d}{ \gamma_L} e^{- k \gamma_L L}, 
\end{equation}
and
\begin{equation} \label{eq:itbd22}
 \sup_{\theta \in [0, 2\pi]} \int_0^{\infty} \, \int_0^{ \infty} \,
 \left| \, \langle A \alpha_{i(s+t)} \left( B (\theta) \right) \rangle_{\theta}  
\, \right| \, ds \, dt   \, \leq \, \, C \, \| A \| \, |X| \,
\frac{L^{d+1}}{ \gamma_L^2} e^{- k \gamma_L L}, 
\end{equation}
respectively.
\end{Remark}

For the next lemma we will need the following basic estimate involving the
decay of certain Fourier transforms. 

\begin{prop} \label{prop:ftbds} Let $a >0$ and $T >0$ be given. Define a function
  $F_{a, T} : \mathbb{R} \to \mathbb{C}$ by 
\begin{equation} \label{eq:fatedef}
F_{a,T}(E) \, := \, \frac{1}{2 \pi i} \, \int_0^T \, e^{- a
  t^2} \, \int_{- \infty}^{\infty} \, \frac{e^{-iEs} \, e^{- a
    s^2}}{s-it} \, ds \, dt. 
\end{equation}
For all $E \in \mathbb{R}$, $F_{a,T}(E) \geq 0$ and the estimate
\begin{equation} \label{eq:FbdE+}
 F_{a,T}(E)  \, \leq \, \frac{T}{2} \, e^{ - \frac{E^2}{4
    a}},
\end{equation}
is valid for $E \geq 0$. In the parameter range, $E \geq 2
a T > 0$, one may also show that
\begin{equation} \label{eq:FbdE-}
\int_0^T \, e^{-Et} \, dt \, - \,  F_{a,T}(-E)   \,
\leq \, \frac{T}{2} \, e^{ - \frac{ E^2}{4 a}}.
\end{equation}
\end{prop}
\begin{proof}
One may easily  verify that for any $t>0$,
\begin{equation} \label{eq:ftstuff}
\frac{1}{2 \pi i} \, \int_{- \infty}^{ \infty}
   \, \frac{ e^{-i E s} \, e^{ - a s^2}}{s -
    i t} \, ds \, = \, \frac{1}{2 \sqrt{ \pi a}} \, \int_{0}^{ \infty}
   \, e^{-tw} \, e^{- \frac{(w+E)^2}{4 a}} \, dw,
\end{equation}
for all $E \in \mathbb{R}$, see e.g. Lemma~1 in
\cite{nachtergaele2006a}. This implies the first claim. Evaluating 
the Gaussian integral yields
\begin{equation}
 \frac{1}{2 \sqrt{ \pi a}} \, \int_{0}^{ \infty}
   \, e^{-tw} \, e^{- \frac{(w+E)^2}{4 a}} \, dw \, \leq \,
   \frac{1}{2} \, e^{- \frac{E^2}{4 a}},
\end{equation}
in the case that $E \geq 0$, from which (\ref{eq:FbdE+}) is clear.

To obtain (\ref{eq:FbdE-}), we first recall that the Fourier transform
of a Gaussian is a Gaussian, i.e., for all $z \in \mathbb{C}$,
\begin{equation}
e^{- \frac{z^2}{2}} \, = \, \frac{1}{ \sqrt{2 \pi}} \, \int_{-
  \infty}^{\infty} \, e^{- \frac{x^2}{2}} \, e^{- i x z} \, dx,
\end{equation}
and therefore, by rescaling $z \mapsto - \sqrt{2 a} z$,
multiplying through by $e^{iEz}$ (for $E \in \mathbb{R}$), and
changing variables $w = \sqrt{2 a}x + E$, we have that
\begin{equation} \label{eq:ftg}
e^{iEz} \, e^{- a z^2} \, = \, \frac{1}{2 \sqrt{ \pi a}} \,
\int_{- \infty}^{\infty} \, e^{iwz} \, e^{- \frac{(w-E)^2}{4 a}}
\, dw,
\end{equation}
for all $z \in \mathbb{C}$.

Now, by direct substitution into (\ref{eq:ftstuff}), we have that
\begin{equation}
F_{a,T}(-E) \, = \, \frac{1}{2 \sqrt{ \pi a}} \, \int_0^T \,
e^{- a t^2} \, \int_{0}^{ \infty}
   \, e^{-tw} \, e^{- \frac{(w-E)^2}{4 a}} \, dw.
\end{equation}
Applying (\ref{eq:ftg}), with the special choice of $z=it$, one sees
that
\begin{equation}
\int_0^T e^{-Et} \, dt \, - \, F_{a,T}(-E) \,  = \, \frac{1}{2 \sqrt{ \pi a}} \, \int_0^T \,
e^{- a t^2} \, \int^{0}_{- \infty}
   \, e^{-tw} \, e^{- \frac{(w-E)^2}{4 a}} \, dw.
\end{equation}
Since $w<0$ and $t>0$, the integrand above
\begin{equation}
e^{-tw} \, e^{- \frac{(w-E)^2}{4 a}} \, = \, e^{- \frac{E^2}{4
    a}} \, e^{ \frac{(E-2 a t)w}{2 a}} \, e^{-
  \frac{w^2}{4 a}}
\end{equation}
satisfies a trivial bound when $E \geq 2 a T$. For these values
of $E$, (\ref{eq:FbdE-}) holds.
\end{proof}

We may now prove the main estimate for gapped systems. Recall the definitions
of the operators $B=B(A,H)$ and $B_{a,T}=B_{a,T}(A,H)$ in (\ref{eq:b0})
and (\ref{eq:batgen}), respectively.

\begin{lem} \label{lem:gapbd} Let $H \geq 0$ be a self-adjoint
  operator and $P_E$ denote the family of spectral projections 
  corresponding to $H$. 
  Suppose $H$ has a gap $\gamma >0$, and let $A$ be a local observable
  for which $P_0 A P_0 = 0$. If $2 a T \leq \gamma$, then one has that 
\begin{equation} \label{eq:normdif}
\left\| \, \left( B_{a,T} - B \right) \, P_0 \, \right\| \, \leq \,
 T e^{- \frac{ \gamma^2}{4 a}} \, \left( \frac{ \| A P_0 \| + \|A^*
     P_0 \|}{2} \right) \, + \, \frac{e^{- \gamma
      T}}{ \gamma} \, \| A P_0 \|.
\end{equation}
\end{lem}
\begin{proof}
One may rewrite the difference in these operators as 
\begin{eqnarray}
\left( B_{a, T} - B \right) \, P_0 & = & \int_0^T \left(
  \alpha_{it} \left( A \right) - A_{a}(it) \right) dt \, P_0 
 \nonumber \\
& \mbox{ } & \, + \, \int_T^{\infty} \alpha_{it}
\left( A \right) dt  \, P_0 \, + \, \int_0^T
A_{a}(it)^{*} dt \, P_0 ,
\end{eqnarray}
Each of these terms may be bounded in norm.

For any vectors $f$ and $g$, one may calculate
\begin{eqnarray}
 \langle \, f, \, \int_T^{ \infty} \alpha_{it}(A) dt P_0  \, g \, \rangle
& = &  \int_T^{ \infty} \, \langle \, f, e^{-tH} \, A P_0 g \, \rangle
\, dt \\ & = &  \int_T^{ \infty} \, \int_{\gamma}^{\infty} e^{-tE} \, 
d \langle \, f, P_E\, A P_0 g \, \rangle \, dt, \nonumber 
\end{eqnarray}
where we have used the spectral theorem to rewrite the time evolution
and the fact that $P_0AP_0 = 0$. Clearly then, 
\begin{equation}
\left| \langle \, f, \, \int_T^{ \infty} \alpha_{it}(A) dt P_0   \, g
  \, \rangle \right| \, \leq \, \| f \| \, \| A P_0 g \| \, \int_T^{
  \infty} e^{- \gamma t} \, dt, 
\end{equation}
and therefore,
\begin{equation}
\left\| \int_T^{ \infty} \, \alpha_{it}(A) \, dt \,P_0 \, \right\| \,
\leq \, \frac{e^{- \gamma T}}{ \gamma} \, \| A P_0 \|.
\end{equation}

Likewise, one may similarly calculate
\begin{eqnarray}
\langle f, \, \int_0^T A_{a}(it)^* dt \, P_0 g \, \rangle
& = &  - \int_0^T \, \frac{e^{- a t^2}}{2 \pi i} \, \int_{-
  \infty}^{\infty} \, \langle \, \alpha_s(A) f, P_0 g \, \rangle
\, \frac{e^{- a s^2}}{s+it} \, ds \, dt \nonumber \\ 
& = &  \int_{\gamma}^{ \infty} \, F_{a,T}(E) \, 
d \langle \, f, P_E \, A^* P_0 g \, \rangle ,  
\end{eqnarray}
where we have introduced $F_{a,T}(E) = \overline{F_{a,T}(E)}$ as in
(\ref{eq:fatedef}) of Proposition~\ref{prop:ftbds} above. The estimate
\begin{equation}
\left\| \int_0^T \, A_{a}(it)^{*} \, dt \,P_0 \, \right\| \,
\leq \,  \frac{T}{2} \, e^{ - \frac{\gamma^2}{4 a}} \, \| A^* P_0 \|,
\end{equation}
readily follows from (\ref{eq:FbdE+}) of Propostion~\ref{prop:ftbds}
and the fact that $0 < \gamma \leq E$.

Lastly, an analogous calculation shows that
\begin{eqnarray}
\int_0^T \, \langle \, f, \left[  \alpha_{it}(A) -   A_{a}(it)
\right] \, P_0 g \, \rangle \, dt  \, & =  \quad \quad \quad \nonumber \\
 \quad \quad \quad \int_{\gamma}^{ \infty} \, \left[ \, \int_0^T e^{-Et} \, dt \, -
  F_{a,T}(-E) \, \right] \, d \langle \, f, P_E \, A P_0 g \, \rangle \, .
\end{eqnarray}
Thus, for $2 a T \leq \gamma$, we may apply (\ref{eq:FbdE-}) of 
Proposition~\ref{prop:ftbds} and establish the bound
\begin{equation}
\left\|  \, \int_0^T  \, \left[ \alpha_{it}(A) - A_{a}(
    it) \, \right] \, dt \, P_0 \, \right\| \, \leq \, \frac{T}{2} \, e^{- \frac{ \gamma^2}{4
    a}} \, \| A P_0 \|.
\end{equation}
 
Compiling our estimates, we have proven that: if $2 a T \leq
\gamma$, then
\begin{equation}
\left\| \, \left( B_{a,T} - B \right) \, P_0 \, \right\| \, \leq \,
 T e^{- \frac{ \gamma^2}{4 a}} \, \left( \frac{ \| A P_0 \| + \|A^*
     P_0 \|}{2} \right) \, + \, \frac{e^{- \gamma
      T}}{ \gamma} \, \| A P_0 \|,
\end{equation}
as claimed.
\end{proof}

\begin{Remark} \label{rem:projest} Applying Lemma~\ref{lem:gapbd} to the
  operator $H = H_{\theta, -\theta}$, whose spectral projections we
  denote by $P_E^{\theta}$, and the local observable $A =
  A_1(\theta)$, we find that there exists a constant $C>0$ for which,
  along the parametrization $a = \gamma_L / L$ and $T = L /2$, 
\begin{equation}
\sup_{\theta \in [0, 2\pi] } \left\| \, \left( B_{a,T}(\theta) - B_1(\theta) \right) \, P_0^{\theta} \, \right\| \, \leq \,
C \, L^d \, e^{- \frac{\gamma_L L}{4}} \left( 1 + \frac{ e^{- \frac{\gamma_L L}{4}}}{ \gamma_L L} \right) .
\end{equation}
\end{Remark}



\subsection{Norm preserving flows}\label{sec:npf}
In this section, we collect some basic facts about the solutions of first
order, inhomogeneous differential equations. 
\begin{Definition}
Let $\mathcal{B}$ be a Banach space. For each $\theta \in \R$, 
let $A( \theta) : \mathcal{B} \to \mathcal{B}$ be a bounded linear
operator, and denote by $X( \theta)$ the solution of the 
differential equation 
\begin{equation} \label{eq:fode}
\partial_{\theta} X( \theta) \, = \, A( \theta) \, X( \theta)
\end{equation}
with boundary condition $X(0) = X_0 \in \mathcal{B}$. We say that the
family of operators $A(\theta)$ is {\em norm-preserving} if the corresponding
flow is isometric, i.e., for 
every $X_0 \in \mathcal{B}$, the mapping $\gamma_{\theta} :
\mathcal{B} \to \mathcal{B}$ which associates $X_0 \to X( \theta)$,
i.e., $\gamma_{\theta}(X_0) = X( \theta)$, satisfies
\begin{equation} \label{eq:normp}
\| \, \gamma_{\theta}(X_0) \, \| \, = \, \| \, X_0 \, \| \quad \mbox{for all } \theta \in
\R.
\end{equation}
\end{Definition}

Two typical examples are the case where $\mathcal{B}$ is a Hilbert space 
and $A(\theta)$ is anti-hermitian and the case where $\mathcal{B}$ is a
Banach space of linear operators on a Hilbert space with a spectral
norm (such as a $p-$norm with $p\in[1,+\infty]$), and where
$A(\theta)$ is a symmetric derivation (e.g., $i$ times the commutator with
a self-adjoint operator). 

\begin{thm} \label{thm:normp} Let $A( \theta)$, for $\theta \in \R$, be a family of 
norm preserving opeartors in some Banach space $\mathcal{B}$. For any
bounded measurable function $B : \R \to \mathcal{B}$, the solution of 
\begin{equation} \label{eq:inhom}
\partial_{\theta} Y( \theta) \, = \, A( \theta) Y( \theta) \, + \, B( \theta),
\end{equation}
with boundary condition $Y(0) = Y_0$, satisfies the bound
\begin{equation} \label{eq:yest}
\| \, Y( \theta) \, - \, \gamma_{\theta}(Y_0) \, \| \, \leq \, \int_0^{ \theta}  \| \, B( \theta')
\, \| \, d \theta' .
\end{equation}
\end{thm}

\begin{proof}
For any $\theta \in \R$, let $X( \theta)$ be the solution of 
\begin{equation} \label{eq:fode1}
\partial_{\theta} X( \theta) \, = \, A( \theta) \, X( \theta)
\end{equation}
with boundary condition $X(0) = X_0$, and let $\gamma_{\theta}$ be the
linear mapping which takes $X_0$ to $X( \theta)$. By variation of constants,
the solution of the inhomogeneous equation (\ref{eq:inhom}) may be
expressed as
\begin{equation} \label{eq:ysol}
Y( \theta) \, = \, \gamma_{\theta} \left( \, Y_0 \, + \, \int_0^{
      \theta} ( \gamma_s)^{-1} \left( B(s) \right) ds \, \right).
\end{equation}
The estimate (\ref{eq:yest}) follows from (\ref{eq:ysol}) as $A( \theta)$ is
norm preserving.
\end{proof}

\subsection{Existence of local unitaries with vanishing expectation}\label{sec:zerounitary}

Consider a finite system with a Hamiltonian of the form
\begin{equation}
H=\sum_{X}\Phi(X)
\end{equation}
where $\Phi$ is an interaction as defined at the beginning of the paragraph containing
equation (\ref{eq:defnphia}).  In the introduction, (\ref{eq:ugb}), we stated a simple upper 
bound for the spectral gap of any such Hamiltonian with a unique ground state.
The argument we gave there made use of a one-site unitary $U\in\mathcal{A}_{\{x\}}$
with the property that $\langle \Omega, U\Omega\rangle = 0$. In the following lemma
we show that such a unitary always exists.

\begin{lem}\label{lem:zerounitary}
Let $\mathcal{H}$ be a complex Hilbert space of dimension at least $2$. Then, for any
density matrix $\rho$ on $\mathcal{H}$, there exists a unitary $U\in\mathcal{B}(\mathcal{H})$
such that $\Tr \rho U =0$.
\end{lem}
\begin{proof}
First consider the case where $\dim\mathcal{H}$ is finite and even, or infinite. Let
$\{e_0,e_1,\ldots\}$ denote an orthonormal basis of eigenvectors of $\rho$, with eigenvalues
$\rho_i$ ordered in non-increasing order. If $\mathcal{H}$ is
not separable, it is sufficient that $\{e_0,e_1,\ldots\}$ contain a basis for the separable
subspace $\ran \rho$.) Then, a suitable unitary $U$ can be defined as follows:
\begin{equation}\label{eq:defzerounitary}
U=\bigoplus_{i\geq 0} \ketbra{e_{2i+1}}{e_{2i}} + \ketbra{e_{2i}}{e_{2i+1}}\, .
\end{equation}
If $\dim \mathcal{H}$ is odd (and hence by our assumptions $\geq 3$), it is
sufficient to change the first summand in (\ref{eq:defzerounitary}) as follows
\begin{eqnarray*}
U &=& a\ketbra{e_0}{e_0}-\overline{a}\ketbra{e_1}{e_1}
+b \ketbra{e_1}{e_0} + \overline{b} \ketbra{e_0}{e_1}
+e^{i\phi}\ketbra{e_2}{e_2}\\
&&+ \bigoplus_{i\geq 2} \ketbra{e_{2i-1}}{e_{2i}} + \ketbra{e_{2i}}{e_{2i-1}}\, ,
\end{eqnarray*}
where $a,b\in \mathbb{C}$ and $\phi\in\mathbb{R}$, are chosen such
that $\vert a\vert^2 + \vert b\vert^2 =1$ and
$$
e^{i\phi}\rho_2=a\rho_0 -\overline{a}\rho_1\,.
$$
This is always possible since $\rho_2^2\leq (\rho_0+\rho_1)^2$.

It is straightforward to check that $U$ thus defined has the desired properties.
\end{proof}

\subsection*{Acknowledgement}
         
This article is based on work supported by the U.S. National Science
Foundation under Grant \# DMS-06-05342. Both authors thank the Erwin 
Schr\"odinger Institute for kind hospitality during a fruitful stay.


\begin{thebibliography}{10}

\bibitem{affleck1986}
I.~Affleck and E.H. Lieb, \emph{A proof of part of {H}aldane's conjecture on
  quantum spin chains}, Lett. Math. Phys. \textbf{12} (1986), 57--69.

\bibitem{affleck1988}
I.~Affleck, T.~Kennedy, E.H.~Lieb and H.~Tasaki,
\emph{Valence Bond Ground States in Isotropic Quantum Antiferromagnets},
Commun. Math. Phys. \textbf{115} (1988), 477--528.

\bibitem{aizenman1994}
M.~Aizenman and B.~Nachtergaele, \emph{Geometric aspects of quantum spin
  states}, Comm. Math. Phys. \textbf{164} (1994), 17--63.

\bibitem{bratteli1997}
O.~Bratteli and D.~Robinson, \emph{Operator algebras and quantum statistical
  mechanics 2}, second ed., Springer Verlag, New York, NY, 1997.

\bibitem{Dyson1978}
F.~Dyson, E.H.~Lieb and B.~Simon,
\emph{Phase transitions in quantum spin systems with isotropic and 
non-isotropic interactions},  J. Stat. Phys. \textbf{18} (1978), 335--383.

\bibitem{Haldane1983a}
F.D.M.~Haldane, \emph{Continuum dynamics of the 1-D Heisenberg antiferromagnet:
Identification with the $O(3)$ nonlinear sigma model},  
Phys. Lett. \textbf{93A} (1983), 464--468.

\bibitem{Haldane1983b}
F.D.M.~Haldane, \emph{Nonlinear field theory of large-spin Heisenberg 
antiferromagnets: Semiclassically quantized solitons of the one-dimensional 
easy-axis N\'eel state}, Phys. Rev. Lett. \textbf{50} (1983), 1153--1156.

\bibitem{hastings2004}
M.B. Hastings, \emph{Lieb-{S}chultz-{M}attis in higher dimensions}, Phys.\
  Rev.\ B \textbf{69} (2004), 104431--14.

\bibitem{hastings2006}
M.B. Hastings and T.~Koma, \emph{Spectral gap and exponential decay of
  correlations}, Commun. Math. Phys. \textbf{265} (2006), 781--804,
  \mbox{arXiv:math-ph/0507008}.

\bibitem{Kennedy1988}
T.~Kennedy, E.H.~Lieb, B.S.~Shastry, \emph{Existence of N\'eel order in some 
spin 1/2 Heisenberg antiferromagnets},  J. Stat. Phys. \textbf{53} (1988), 
1019--1030. 
 
\bibitem{Koma1994}
T.~Koma and H.~Tasaki, \emph{Symmetry breaking and finite-size effects 
in quantum many-body systems},  J. Stat. Phys. \textbf{76} (1994), 745--803.

\bibitem{Landau1981}
L.~Landau, J.~Fernando-Perez, W.F.~Wreszinski, \emph{Energy gap, clustering, 
and the Goldstone theorem in statistical mechanics},  J. Stat. Phys.
\textbf{26} (1981), 755--766. 

\bibitem{Lieb1962}
E.~Lieb and D.~Mattis, \emph{Ordering energy levels in interacting spin
  chains}, Journ. Math. Phys. \textbf{3} (1962), 749--751.

\bibitem{Lieb1961}
E.~Lieb, T.~Schultz, and D.~Mattis, \emph{Two soluble models of an
  antiferromagnetic chain}, Ann.~Phys.~(N.Y.) \textbf{16} (1961), 407--466.

\bibitem{Lieb1972}
E.H. Lieb and D.W. Robinson, \emph{The finite group velocity of quantum spin
  systems}, Comm. Math. Phys. \textbf{28} (1972), 251--257.

\bibitem{Manousakis1991}
E.~Manousakis, \emph{The spin-$1/2$ {H}eisenberg antiferromagnet on a square
  lattice and its application to the cuprous oxides}, Rev. Mod. Phys.
  \textbf{63} (1991), 1--62.

\bibitem{Nachtergaele1994}
B.~Nachtergaele, \emph{Quasi-state decompositions for quantum spin systems},
  Probability Theory and Mathematical Statistics. Proceedings of the 6th
  Vilnius Conference (et.~al. Grigelionis, B., ed.), Utrecht-Tokyo-Vilnius,
  VSP/Tev, 1994, \mbox{cond-mat/9312012cond-mat/9312012}, pp.~565--590.

\bibitem{nachtergaele2006b}
B.~Nachtergaele, Y.~Ogata, and R.~Sims, \emph{Propagation of correlations in
  quantum lattice systems}, \mbox{arXiv:math-ph/0603064}, to appear in J. Stat.
  Phys. (2006).

\bibitem{nachtergaele2006c}
B.~Nachtergaele and R.~Sims, \emph{Recent progress in quantum spin systems},
  Markov Processes Rel. Fields, \mbox{arXiv:math-ph/0512020}, to appear (J.T.
  Lewis special edition).

\bibitem{nachtergaele2006a}
\bysame, \emph{Lieb-{R}obinson bounds and the exponential clustering theorem},
  Commun. Math. Phys. \textbf{265} (2006), 119--130,
  \mbox{arXiv:math-ph/0506030}.

\bibitem{Sachdev1995}
S.~Sachdev, \emph{Quantum antiferromagnets in two dimensions}, Low dimensional
  quantum field theories for condensed matter physicists (Singapore) (Yu~Lu,
  S.~Lundqvist, and G.~Morandi, eds.), World Scientific, 1995,
  \mbox{arXiv:cond-mat/9303014}.

\bibitem{Sachdev2002}
S.~Sachdev and K.~Park, \emph{Ground states of quantum antiferromagnets in two
  dimensions}, Annals of Physics (N.Y.) \textbf{298} (2002), 58--122.

\bibitem{Simon1993}
B.~Simon, \emph{The statistical mechanics of lattice gases}, Princeton Series
  in Physics, vol.~1, Princeton University Press, Princeton, NJ, 1993.

\bibitem{tasaki2004}
H.~Tasaki, \emph{Low-lying excitation in one-dimensional lattice electron
  system}, \mbox{arXiv:cond-mat/0407616}.

\bibitem{yamanaka1997}
M.~Yamanaka, M.~Oshikawa, and I.~Affleck, \emph{Nonperturbative approach to
  {L}uttinger's theorem in one dimension}, Phys. Rev. Lett. \textbf{79} (1997),
  1110--1113.

\end{thebibliography}
\providecommand{\bysame}{\leavevmode\hbox to3em{\hrulefill}\thinspace}

\end{document}